\documentclass[11pt]{article}			
\usepackage{aaspp4}

\newcommand{\kms}{km~s$^{-1}$}
\newcommand{\ergss}{ergs~s$^{-1}$}

\newcommand{\Ha}{H$\alpha$}
\newcommand{\Hb}{H$\beta$}
\newcommand{\HI}{\ion{H}{1}}	
\newcommand{\OII}{\ion{O}{2}}	
\newcommand{\NII}{\ion{N}{2}}	
\newcommand{\SII}{\ion{S}{2}}	
\newcommand{\Msun}{$M_{\sun}$}
\newcommand{\Zsun}{$Z_{\sun}$}
\newcommand{\HST}{{\it HST\/}}
\newcommand{\eg}{e.g., }
\newcommand{\ie}{i.e., }
\newcommand{\cf}{cf.\ }

\newcounter{tmpii}\newlength{\tmplngth}

\catcode`\@=11
\let\@internalcite\cite
\def\cite{\def\@citeseppen{-1000}\def\@cite##1##2{(##1\if@tempswa , ##2\fi)}\def\citeauthoryear##1##2##3{##1 ##3}\@internalcite}

\def\citeNP{\def\@citeseppen{-1000}\def\@cite##1##2{##1\if@tempswa , ##2\fi}\def\citeauthoryear##1##2##3{##1 ##3}\@internalcite}
\def\citeN{\def\@citeseppen{-1000}\def\@cite##1##2{##1\if@tempswa , ##2)\else{)}\fi}\def\citeauthoryear##1##2##3{##1 (##3}\@citedata}
\def\citeA{\def\@citeseppen{-1000}\def\@cite##1##2{(##1\if@tempswa , ##2\fi)}\def\citeauthoryear##1##2##3{##1}\@internalcite}
\def\citeANP{\def\@citeseppen{-1000}\def\@cite##1##2{##1\if@tempswa , ##2\fi}\def\citeauthoryear##1##2##3{##1}\@internalcite}
\def\shortcite{\def\@citeseppen{-1000}\def\@cite##1##2{(##1\if@tempswa , ##2\fi)}\def\citeauthoryear##1##2##3{##2 ##3}\@internalcite}
\def\shortciteNP{\def\@citeseppen{-1000}\def\@cite##1##2{##1\if@tempswa , ##2\fi}\def\citeauthoryear##1##2##3{##2 ##3}\@internalcite}
\def\shortciteN{\def\@citeseppen{-1000}\def\@cite##1##2{##1\if@tempswa , ##2)\else{)}\fi}\def\citeauthoryear##1##2##3{##2 (##3}\@citedata}
\def\shortciteA{\def\@citeseppen{-1000}\def\@cite##1##2{(##1\if@tempswa , ##2\fi)}\def\citeauthoryear##1##2##3{##2}\@internalcite}
\def\shortciteANP{\def\@citeseppen{-1000}\def\@cite##1##2{##1\if@tempswa , ##2\fi}\def\citeauthoryear##1##2##3{##2}\@internalcite}
\def\citeyear{\def\@citeseppen{-1000}\def\@cite##1##2{(##1\if@tempswa , ##2\fi)}\def\citeauthoryear##1##2##3{##3}\@citedata}
\def\citeyearNP{\def\@citeseppen{-1000}\def\@cite##1##2{##1\if@tempswa , ##2\fi}\def\citeauthoryear##1##2##3{##3}\@citedata}
\def\@citedata{\@ifnextchar [{\@tempswatrue\@citedatax}{\@tempswafalse\@citedatax[]}}
\def\@citedatax[#1]#2{%
\if@filesw\immediate\write\@auxout{\string\citation{#2}}\fi%
  \def\@citea{}\@cite{\@for\@citeb:=#2\do%
    {\@citea\def\@citea{, }\@ifundefined%
       {b@\@citeb}{{\bf ?}%
       \@warning{Citation `\@citeb' on page \thepage \space undefined}}%
{\csname b@\@citeb\endcsname}}}{#1}}%
\def\@citex[#1]#2{%
\if@filesw\immediate\write\@auxout{\string\citation{#2}}\fi%
  \def\@citea{}\@cite{\@for\@citeb:=#2\do%
    {\@citea\def\@citea{; }\@ifundefined%
       {b@\@citeb}{{\bf ?}%
       \@warning{Citation `\@citeb' on page \thepage \space undefined}}%
{\csname b@\@citeb\endcsname}}}{#1}}%

\def\@biblabel#1{}
\newlength{\bibhang}
\catcode`\@=12

\begin{document}

\lefthead{Koekemoer et al.}
\righthead{The Blue Continuum, Line Emission and Radio Galaxy in Abell~2597}

\hbox{}\vspace{-60pt}
\title{The Extended Blue Continuum and Line Emission around the Central Radio
	Galaxy in Abell~2597%
\footnote{Based on observations made with the NASA/ESA Hubble Space Telescope,
obtained at the Space Telescope Science Institute, which is operated by the
Association of Universities for Research in Astronomy, Inc., under NASA
contract NAS~5-26555.}
}

\author{Anton~M.~Koekemoer\altaffilmark{2,3},
	Christopher~P.~O'Dea\altaffilmark{2},
	Craig~L.~Sarazin\altaffilmark{4},
	Brian~R.~McNamara\altaffilmark{5},
	Megan~Donahue\altaffilmark{2},	
	G.~Mark~Voit\altaffilmark{2},
	Stefi~A.~Baum\altaffilmark{2},
	Jack~F.~Gallimore\altaffilmark{6}}

\altaffiltext{2}{Space Telescope Science Institute,
	3700 San Martin Drive, Baltimore, MD 21218;\\
	koekemoe@stsci.edu, odea@stsci.edu, donahue@stsci.edu, voit@stsci.edu,
	sbaum@stsci.edu}

\altaffiltext{3}{NASA / Goddard Space Flight Center,
	Code 630, Greenbelt, MD 20771;\\
	koekemoer@gsfc.nasa.gov}

\altaffiltext{4}{\raggedright
	Department of Astronomy, University of Virginia,
	P.O. Box 3818, Charlottesville, VA 22903-0818;
	cls7i@coma.astro.virginia.edu}

\altaffiltext{5}{\raggedright
	Harvard-Smithsonian Center for Astrophysics,
	60 Garden Street, Cambridge, MA 02138;
	\mbox{brm@head-cfa.harvard.edu}}

\altaffiltext{6}{\raggedright
	National Radio Astronomy Observatory,
	520 Edgemont Road, Charlottesville, VA 22903-2475;
	\mbox{jgallimo@nrao.edu}}

{\centering Submitted to {\it The Astrophysical Journal}\par}

\accepted{1999 June 21}
\sluginfo

\begin{abstract}
We present results from detailed imaging of the centrally dominant radio
elliptical galaxy in the cooling flow cluster Abell~2597, using data obtained
with the Wide Field and Planetary Camera 2 (WFPC2) on the {\it Hubble Space
Telescope (HST)\/}. This object is one of the archetypal ``blue-lobed'' cooling
flow radio elliptical galaxies, also displaying a luminous emission-line
nebula, a compact radio source, and a significant dust lane and evidence of
molecular gas in its center. We show that the radio source is surrounded by a
complex network of emission-line filaments, some of which display a close
spatial association with the outer boundary of the radio lobes. We present a
detailed analysis of the physical properties of ionized and neutral gas
associated with the radio lobes, and show that their properties are strongly
suggestive of direct interactions between the radio plasma and ambient gas. We
resolve the blue continuum emission into a series of knots and clumps, and
present evidence that these are most likely due to regions of recent star
formation. We investigate several possible triggering mechanisms for the star
formation, including direct interactions with the radio source, filaments
condensing from the cooling flow, or the result of an interaction with a
gas-rich galaxy, which may also have been responsible for fueling the active
nucleus. We propose that the properties of the source are plausibly explained
in terms of accretion of gas by the cD during an interaction with a gas-rich
galaxy, which combined with the fact that this object is located at the center
of a dense, high-pressure ICM can account for the high rates of star formation
and the strong confinement of the radio source.
\end{abstract}

\keywords{cooling flows ---
	galaxies: clusters: individual (A2597) ---
	galaxies: jets ---
	galaxies: elliptical and lenticular, cD ---
	galaxies: stellar content ---
	galaxies: individual (PKS~2322$-$123)}

\section{Introduction}

Rich clusters of galaxies generally contain a diffuse intracluster medium (ICM)
having temperatures $T\sim 10^7 - 10^8$~K and densities
$n \sim 10^{-2} - 10^{-4}$~cm$^{-3}$, as inferred from observations at X-ray
and other wavebands
	(\eg \citeNP{Sarazin.1986.RevModPh.58.1};
	\citeNP{Fabian.1994.ARAA.32.277}).
An important question related to the overall physics of the ICM concerns the
central regions of clusters ($r \lesssim 10 - 100$~kpc), where the inferred
ICM densities and pressures in some cases are sufficiently high that cooling to
$T \lesssim 10^4$~K can occur on timescales shorter than the
cluster lifetime
	(\eg \citeNP{Cowie.1977.ApJ.215.723};
	\citeNP{Fabian.1977.MNRAS.180.479}).
These ``cooling flow'' clusters often exhibit intense optical emission-line
nebulae associated with the centrally dominant (cD) galaxies at their centers,
together with blue continuum excess emission, and the strength of these effects
appears to correlate with the cooling rate or central pressure of the X-ray
emitting gas
	(\citeNP{Heckman.1981.ApJ.250.L59};
	\citeNP{Hu.1985.ApJS.59.447};
	\citeNP{Johnstone.1987.MNRAS.224.75};
	\citeNP{Romanishin.1987.ApJ.323.L113};
	\citeNP{Romanishin.1988.ApJ.324.L17};
	\citeNP{Heckman.1989.ApJ.338.48};
	\citeNP{McNamara.1989.AJ.98.2018},
	\citeyearNP{McNamara.1992.ApJ.393.579},
	\citeyearNP{McNamara.1993.AJ.105.417};
	\citeNP{Crawford.1992.MNRAS.259.265},
	\citeyearNP{Crawford.1993.MNRAS.265.431};
	\citeNP{Allen.1995.MNRAS.276.947};
	\citeNP{McNamara.1997.GCCF.109}).
The majority of cooling flow cluster cD galaxies are also radio-loud, hosting
moderately powerful but relatively compact radio sources
	\cite{Burns.1990.AJ.99.14}.

Several intriguing questions have been posed concerning cDs in cooling flows.
For example, one question concerns the energy source for the line emission: the
optical line luminosities are generally one to three orders of magnitude higher
than would be expected from the X-ray-derived cooling rates, if each proton in
the cooling gas experienced only a single recombination
	(\eg \shortciteNP{Heckman.1989.ApJ.338.48};
	\citeNP{Voit.1997.ApJ.486.242}).
This has led to the consideration of numerous other potential sources of
heating, including photoionization by young stars or the AGN, auto-ionizing
shocks, photoionization by X-rays from the ICM, or cosmic-ray heating
	(\eg \citeNP{Voit.1997.ApJ.486.242}
	and references therein).
Likewise the blue continuum emission is of considerable interest, and suggested
possibilities for its origin have included scattered nuclear light, recent star
formation, or inverse Compton emission. Finally, a crucial point to investigate
concerns the inter-relationships between these various components: the cD
elliptical, the emission-line gas, the blue continuum, the radio source, and
the surrounding dense high-pressure ICM: how are these related to one another,
and do they have underlying common physical relationships?

We present results from a detailed study of a comparatively nearby strong
cooling flow cluster, Abell 2597, using images obtained with the Wide Field and
Planetary Camera 2 (WFPC2) on board the {\it Hubble Space Telescope (HST)\/}.
Previous radio and X-ray images have revealed that the cluster contains a
moderately luminous radio source
($P_{1.4~\rm GHz} \sim 6 \times 10^{25}$~W~Hz$^{-1}$,
	\citeNP{Owen.1995.AJ.109.14};
$z_{\rm em} = 0.0821 \pm 0.0002$,
	\citeNP{Voit.1997.ApJ.486.242};
$z_{\rm abs} = 0.0823 \pm 0.0003$,
	\shortciteNP{Taylor.1999.ApJ.512.L27})
embedded in a luminous X-ray halo with $L_X \sim 10^{45}$~\ergss\
	(\shortciteNP{Crawford.1989.MNRAS.236.277};
	\citeNP{Owen.1992.ApJS.80.501};
	\shortciteNP{Sarazin.1995.ApJ.447.559};
	\citeNP{Sarazin.1997.ApJ.480.203}).
The host galaxy of the radio source is a cD elliptical containing an extended,
luminous emission-line nebula
	(\citeNP{Heckman.1989.ApJ.338.48};
	\citeNP{Smith.1989.ApJ.341.658};
	\citeNP{Crawford.1992.MNRAS.259.265};
	\citeNP{Voit.1997.ApJ.486.242}).
The galaxy is remarkable for its ``blue lobes'' of continuum excess emission
that are approximately co-incident with the radio lobes
	(\citeNP{McNamara.1992.ApJ.393.579},
	\citeyearNP{McNamara.1993.AJ.105.417}).
Spatially extended 21-cm \HI\ absorption has also been detected against the
radio lobes
	\shortcite{ODea.1994.ApJ.436.669}.
Although the radio source is comparatively compact (with a size
$\lesssim 5$~kpc), its power is comparable to that of much larger FR~II
(``classical double'') sources, suggesting that the dense intracluster medium
may play a role in its confinement. The source may be the nearest example of
a compact steep-spectrum source
	\shortcite{ODea.1994.ApJ.436.669}.
Here we present data that fully resolve the blue emission in the lobes and
reveal a complex network of emission-line and continuum filaments surrounding
the central source.

All distance-dependent quantities in this paper have been calculated assuming
$H_0 = 75$~\kms~Mpc$^{-1}$ and $q_0 = 0$.
%
%
%
%
%
%
%
%
%
%

\section{Observations}

We observed A2597 in July 1996 with the \HST\ WFPC2 Camera~\#3 (WF3) through
the F410M filter, with the total exposure time being ``CR-split'' into three
equal, consecutive exposures to facilitate the removal of cosmic rays. This is
effectively a narrow-band image, mostly containing flux from the redshifted
[\OII] $\lambda$3727 emission line. In this paper we also discuss new work that
we have carried out on two archival broad-band images of A2597, taken with the
WFPC2 Planetary Camera (PC1) through the F450W and F702W filters (Program ID
\#6228 in the \HST\ Data Archive; see also
	\shortciteNP{Holtzman.1996.AJ.112.416}).
These images contain substantial continuum flux in addition to the redshifted
[\OII] $\lambda$3727 and \Ha\ + [\NII] $\lambda\lambda$6548,6583 emission
lines, and therefore allow a detailed investigation of the emission-line and
stellar continuum properties of this object, when combined with the F410M
image. In Table~\ref{tab:HST_obsns} we present further details of the
observations and filter characteristics.

\begin{deluxetable}{lcccccl}
\tablewidth{0pt}
\tablecaption{\label{tab:HST_obsns}
	\HST\ WFPC2 Imaging Observations of A2597}
\tablehead{
Date		& Exposure	& Filter & $\bar{\lambda}\;$\tablenotemark{a}
								& $\delta \bar{\lambda}\;$\tablenotemark{a}
										& Detector	& Pixel size\tablenotemark{b}	\nl
		& (s)		&	& (\AA)			& (\AA)	
			&		& (arcsec)
}
\startdata
1996 July 27	& 2100		& F410M	& 4090.1		& \phn146.7			& WF3		& 0$\farcs$0996	\nl
1995 July \phn5	& 2500		& F450W	& 4519.1		& \phn956.6			& PC1		& 0$\farcs$0455	\nl
1995 July \phn5	& 2100		& F702W	& 6867.8		& 1382.5				& PC1		& 0$\farcs$0455
\tablenotetext{a}{The specifications for the mean central filter wavelength
	$\bar{\lambda}$ and filter width $\delta \bar{\lambda}$ are given in
	Table~6.1 of the WFPC2 Instrument Handbook
		\cite{Biretta.1996.WFPC2.STScI},
	where definitions for these quantities are also presented.}
\tablenotetext{b}{Pixel sizes are from
	\protect\shortciteN{Holtzman.1995.PASP.107.156}.}
\enddata
\end{deluxetable}

Recalibration of the data was performed using the IRAF/STSDAS%
\footnote{The Image Reduction and Analysis Facility (IRAF) is distributed by
the National Optical Astronomy Observatories, which is operated by the
Association of Universities for Research in Astronomy, Inc., under contract to
the National Science Foundation. The Space Telescope Science Data Analysis
System (STSDAS) is distributed by the Space Telescope Science Institute.}
pipeline processing software for WFPC2 (see also
	\citeNP{Holtzman.1995.PASP.107.156},
	\citeyearNP{Holtzman.1995.PASP.107.1065}),
using updated calibration reference files where applicable. The removal of hot
pixels with elevated dark currents was investigated using the IRAF/STSDAS task
WARMPIX together with a table of hot pixels nearest in time to the
observations, although it was found that most of these were adequately removed
through the use of the appropriate dark frame.

We verified that no significant alignment shifts were present between the
multiple exposures in each of the three filters, before combining them using
the IRAF/STSDAS task CRREJ which removes most of the cosmic rays as well as some
remaining hot and cold pixels. The cosmic ray and bad pixel removal was checked
by examining the residual difference between the cleaned image and each of the
two original frames, and interpolations were performed across the remaining
defects. Sky subtraction in the broad-band F450W PC1 image was carried out by
obtaining a mean value of the sky background in several empty regions of the
frame, and subtracting this as a constant value from the entire image. The sky
background for the F702W PC1 image was determined by scaling the F450W sky
value using the sky count rates presented for these filters in the \HST\ Data
Handbook. This was necessary since the extended envelope of the elliptical is
more prominent in the F702W image than in the F450W image, thereby precluding
an accurate direct determination of the sky background. The sky count rate
values (0.0071 and 0.0028 e$^-$~s$^{-1}$~pixel$^{-1}$ for the F450W and F702W
images respectively) were sufficiently low that residual large-scale
fluctuations in the images were at levels $\lesssim 0.05$~counts, well within
the Poisson noise limits.

The two broad-band images were transformed to F450W and F702W magnitudes,
$m_{\rm F450W}$ and $m_{\rm F702W}$ respectively, in the \HST\ WFPC2 synthetic
``VEGAMAG'' photometric system, defined in
	\citeN{Holtzman.1995.PASP.107.1065},
making use of the IRAF/STSDAS SYNPHOT package to obtain accurate photometric
parameters for the filters used. Conversions to Johnson {\it B\/} and {\it R\/}
magnitudes were not carried out at this point, since substantial differences
between the WFPC2 filter response functions and those of the Johnson standard
system would introduce photometric errors if significant color variations are
present across the galaxy.

\section{Morphological Properties}

%
%
%
%
%
%
%

In Figure~\ref{fig:f702W} we present the broad-band F702W PC1 image of A2597.
It is clear that the central region of the elliptical is dominated by a complex
network of filaments. Some of the filaments appear clumpy and marginally
resolved, particularly the arcs extending toward the north-east and the south.
There is also some evidence of a dust lane crossing the nucleus, at
approximately the same position angle as the major axis of the elliptical. A
small object, possibly a companion galaxy, is present to the north of the
elliptical, and numerous star clusters are also distributed throughout the
elliptical, some of which may be resolved
	(\cf \shortciteNP{Holtzman.1996.AJ.112.416}).

The filamentary structure dominates the narrow-band emission-line image of the
galaxy, as shown in our F410M WF3 image presented in Figure~\ref{fig:f450W_3cm},
which contains the redshifted [\OII] $\lambda$3727 emission. We superpose
contours from a 3.5~cm VLA image of the radio continuum emission
	\shortcite{Sarazin.1995.ApJ.447.559}.
The radio and optical images were aligned on the basis of the radio core
position, which is assumed to coincide with the broad-band optical nucleus of
the galaxy. Having aligned the images on this basis alone, it is remarkable
that the bright northern filamentary arc follows precisely along the outer
contours of the radio lobe. The rest of the northern radio continuum is also
apparently associated with lower-level filamentary emission. Although there is
less optical filamentation associated with the southern radio lobe, an
emission-line filament is present near the eastern edge of this lobe. Closer to
the center, there is a bright, compact region of emission immediately
south-west of the nucleus, which is co-incident with a radio hotspot as well as
an abrupt change in the orientation of the radio lobe. We also note that some
of the filamentary structures within the galaxy are apparently unassociated
with the radio lobes, particularly the filaments extending away from the
nucleus toward the east, and a more diffuse region surrounding the nucleus on
the west.

\subsection{Separating the Continuum and Emission-Line Features}

One of the interesting questions related to the properties of the filaments
concerns the relative fractions of continuum and emission-line contributions to
their total flux. We have investigated this by comparing the narrow-band F410M
image to the broad-band F450W image, both of which contain the redshifted
[\OII] $\lambda$3727 emission line but contain different amounts of stellar
continuum flux. We used the IRAF/STSDAS package DRIZZLE
	\cite{Fruchter.1997.SPIE.3164.120}
to transform the higher-resolution PC1 F450W image to the coarser pixel scale
of the WF3 F410M image, also taking into account the differences in orientation
angles and geometrical distortions across the chips.

We used the SYNPHOT package to flux-calibrate each of these two images in cgs
units (\ergss~cm$^{-2}$~arcsec$^{-2}$). A fundamental uncertainty inherent in
this calibration is the lack of a detailed knowledge of the spectral energy
distribution across the filter bandpass at each pixel in the image. Two
limiting cases are: (1) monochromatic emission, with all the flux being
emitted at one wavelength (appropriate for pixels dominated by a single
emission line); (2) a flat spectral energy distribution, specifically with
constant flux density per unit wavelength $f_\lambda$. For each of the two
filters, we generated a flux calibration factor for each limiting case using
the SYNPHOT task BANDPAR. The difference between the two cases is negligible
for the F410M filter ($\sim 1.4\%$) but is of order $\sim 12\%$ for the broader
F450W filter. Therefore we flux-calibrated the F410M filter using the
monochromatic limit, while the F450W filter was calibrated with the
flat-spectrum approximation, which is justified by the fact that the flux in
this broader filter is dominated by stellar continuum (based on examining the
spatially integrated ground-based spectra of
	\citeNP{Voit.1997.ApJ.486.242}).
We note however that this approximation may break down for individual pixels
with extremely high [\OII] $\lambda$3727 equivalent widths, introducing a
maximum error of $\sim 12\%$ (for pure line emission), or more likely
$\lesssim 5 - 6\%$ (for comparable proportions of line and continuum emission).

We then subtracted the flux-calibrated F410M image from the geometrically
transformed F450W image to investigate whether the filaments display any
residual continuum emission. As described above, the uncertainties in flux
calibration due to differences in the filter bandpasses are likely to be
$\lesssim 5\%$, therefore this subtraction provides a reliable indication of
the relative contribution of emission-line and continuum emission to the
filaments. We found that the extended filaments and arcs around the radio lobes
disappeared almost entirely in the subtracted image, and therefore can be
considered to consist predominantly of [\OII] $\lambda$3727 line emission. The
filaments nearer the galaxy nucleus, however, display a mixture of continuum
and line emission and are apparently dominated in some regions by excess blue
continuum, which we discuss in more detail in the following section.

\subsection{The Blue Continuum Excess%
\label{sec:blue_cont}}

We have used the broad-band PC1 F450W and F702W images to investigate the
detailed spatial distribution and morphology of excess blue emission in A2597,
which was first discovered in ground-based images by
	\citeN{McNamara.1993.AJ.105.417}
as marginally resolved {\it U $-$ I\/} continuum excess aligned along the radio
axis. We first matched both PC1 images to a common point-spread function (PSF)
by convolving each image with the PSF of the other, having computed the
appropriate PSF for each filter band using the \HST\ Tiny Tim software
	\cite{Krist.1997.TinyTim.STScI}.
The images were then calibrated in the \HST\ VEGAMAG system, after which we
calculated the $m_{\rm F450W} - m_{\rm F702W}$ color distribution of the
galaxy, which we present in Figure~\ref{fig:m450W-m702W}. We have also verified
that smoothing the \HST\ data to the resolution of the ground-based images
reveals a good correspondence between the {\it U\/}-band and F450W blue
continuum. The blue emission in Figure~\ref{fig:m450W-m702W} is resolved in
detail into a number of separate components: (1)~a band of extended, intense
blue emission around the ``base'' of the northern radio lobe, immediately
north-east of the nucleus; (2)~a group of blue point-like features south-west
from the southern radio lobe; (3)~a blue arc that appears to trace the edge of
the southern radio lobe; (4)~diffuse blue emission around the radio lobes, and
specifically to the north-west of the nucleus.

The possibility that the blue continuum is predominantly due to scattered
nuclear light is ruled out on the basis of its low observed polarization
	\shortcite{McNamara.1998.prep};
optical synchrotron or inverse Compton emission are also not feasible since
none of the blue continuum sources correspond to strong radio hotspots. The
most plausible remaining explanation is that the blue continuum is due to young
stars, and in this section we examine various star formation scenarios that may
be responsible for producing the extended blue continuum and the isolated blue
clusters. In \S\ref{sec:implications} we present a more detailed discussion of
the relative plausibility of several possible triggering mechanisms for the star
formation, together with various scenarios for the origin of the gas from which
the stars are formed.

\subsubsection{The Extended and Diffuse Blue Continuum}

Our \HST\ images resolve an intense band of blue continuum to the east and
north of the nucleus, extending around toward the west, thereby covering a
spatial scale of several kiloparsecs (Figure~\ref{fig:m450W-m702W}). Some
less intense blue emission is also associated with the southern radio lobe.
Resolving the blue emission allows a more accurate estimate of its intrinsic
color and luminosity than was previously obtainable from the ground-based
measurements obtained by
	\citeN{McNamara.1993.AJ.105.417},
therefore we use our WFPC2 data to investigate the star formation processes
involved.

The intrinsic color of the blue excess is derived from the residual continuum in
each band after subtracting the underlying elliptical fit, as described in more
detail in \S\ref{sec:ell_fits}. In that section we also discuss the effects of
dust obscuration, which is present in the F702W band as well as F450W. Since
the strongest obscuration is confined to a few small localized patches, we are
still able to discuss the overall properties of the remainder of the blue
continuum excess. Both the F450W and F702W bands are also contaminated by
redshifted emission lines, [\OII] $\lambda$3727 and \Ha+[\NII] respectively.
However, the total integrated flux from the excess blue emission in the F450W
image is substantially higher than the flux from the [\OII] $\lambda$3727
emission in that region observed through the F410M filter; this also applies to
the comparison between the F702W excess continuum and the \Ha+[\NII] flux
scaled from the [\OII] $\lambda$3727 flux using the mean
\Ha+[\NII]~/ [\OII] $\lambda$3727 ratio for this object. Furthermore, the bulk
of the line emission displays little spatial correspondence with the blue
continuum. Therefore in this analysis we neglect to first order the
contribution from line emission to the continuum color; spatially resolved
spectrophotometry would be required to further quantify its contribution. We
measure the intrinsic color of the blue excess to be in the range
$m_{\rm F450W} - m_{\rm F702W} \sim 0.4 - 0.8$ with a surface brightness
ranging from $m_{\rm F450W} \sim 22.0 - 20.8$ mag~arcsec$^{-2}$; the integrated
absolute magnitude of the entire extended region of blue continuum excess is
$M_{\rm F450W} \sim -18.7$ (spread over a projected area of $\sim 20$~kpc$^2$).

\begin{figure}
\plotfiddle{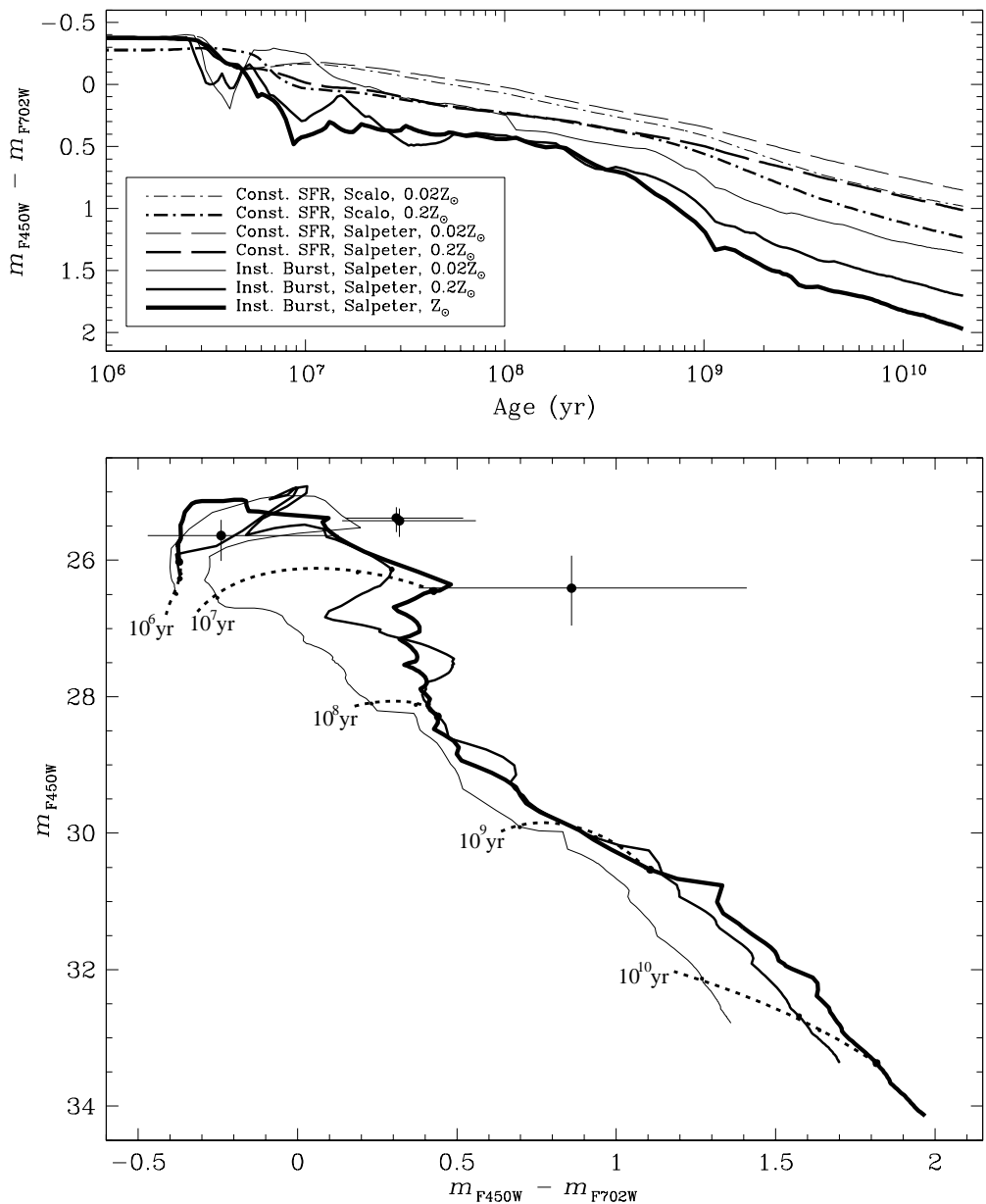}{460pt}{0}{70}{70}{-230pt}{-55pt}
\figcaption[fig4.ps]
{Stellar evolution models
	(\protect\citeNP{Charlot.1999.prep}),
K-corrected to the redshift of A2597 and calibrated for our \HST\ WFPC2 filter
bandpasses. In the top panel we plot the color evolution of models with
constant SFR, for Salpeter and Scalo IMFs, as well as instantaneous burst models
with a Salpeter IMF and metallicities of \Zsun, 0.2\Zsun\ and 0.02\Zsun. For
purposes of clarity, the instantaneous burst models with a Scalo IMF are
omitted from the plot since their color evolution is very similar to the
Salpeter models. The constant SFR models, on the other hand, show little
difference between \Zsun\ and 0.2\Zsun\ models, therefore we plot only
0.2\Zsun\ and 0.02\Zsun\ for these. The lower panel displays the evolution in
color-magnitude space for instantaneous burst models of $10^5$~\Msun\ with a
Salpeter IMF and metallicities of \Zsun, 0.2\Zsun\ and 0.02\Zsun\ (represented
by the same line styles as in the top panel). Again, Scalo models are omitted,
the color differences between the Scalo and Salpeter models being much less
than the formal uncertainties on our photometry. We overplot the measured
colors and apparent magnitudes of the four blue compact objects south-west from
the southern radio lobe. The ages of the model bursts are indicated on the
plot, and the inferred ages of the clusters are discussed in the text.%
\label{fig:charlot_models}}
\end{figure}

We compare the measured properties of the blue continuum to a set of stellar
population synthesis models
	(\citeNP{Charlot.1999.prep}, private communication),
which have been K-corrected to the redshift of A2597 and calibrated in the
\HST\ VEGAMAG system using our \HST\ WFPC2 filter bandpass transmission curves.
We find it preferable to retain our measured magnitudes in the VEGAMAG system
and compare them with models that have also been calculated in this system,
rather than attempting to transform the magnitudes to other systems such as the
Johnson/Cousins system, because uncertainties in the color terms can translate
into errors of several tenths of a magnitude. The models are shown in
Figure~\ref{fig:charlot_models} and represent two scenarios for star formation:
(1)~a constant star formation rate (SFR), assuming an invariant IMF; (2)~an
instantaneous burst of star formation (or ``Simple Stellar Population'', SSP).
In each case we investigated two standard IMFs for which models were available,
namely
	\citeN{Salpeter.1955.ApJ.121.161}
and
	\citeN{Scalo.1986.FCP.11.1},
each with upper and lower mass cutoffs of 0.1 and 100~\Msun\ respectively.
We also examined models of different metallicities, specifically \Zsun,
0.2\Zsun\ and 0.02\Zsun. Further details of the model parameterizations are
described in
	Charlot \& Bruzual (1991, 1999)
	\nocite{Charlot.1999.prep,Charlot.1991.ApJ.367.126}
and
	\citeN{Bruzual.1993.ApJ.405.538}.
Having calculated \HST\ colors and magnitudes for these various models, we
note that the difference between the color evolution of \Zsun\ and 0.2\Zsun\
models is minimal for the constant SFR models, thus in
Figure~\ref{fig:charlot_models} we only plot the 0.2\Zsun\ models for the
purposes of clarity. For the SSP models, the color evolution is more sensitive
to metallicity but is relatively insensitive to the choice of IMF, therefore
the SSP models that we plot are for the Salpeter IMF only.

In Table~\ref{tab:blue_extended} we present the inferred star formation
parameters that are required to match the observed range of colors and
integrated absolute magnitude of the extended blue continuum distribution,
for the various star formation scenarios that we have described. For each
model, we tabulate the upper and lower limits of the range of epochs in its
evolution that reproduce our observed $m_{\rm F450W} - m_{\rm F702W}$ color.
For constant SFR models, the SFR for each of the two epochs is obtained by
matching the model to the $M_{\rm F450W}$ absolute magnitude. For the
instantaneous burst models we tabulate the total mass of stars (an SFR is not
physically meaningful in this case since the burst is represented by a
delta-function).

We point out that the inferred ages are all highly dependent upon the measured
colors, and should therefore be interpreted as {\it upper limits} due to the
likely presence of reddening across most of the field. For example, a decrease
in the intrinsic $m_{\rm F450W} - m_{\rm F702W}$ color by only a further
0.2~magnitudes will yield ages as low as $6 \times 10^7$~yr for the continuous
SFR models, and $\sim 5 \times 10^6$~yr for the instantaneous burst models.
However, due to our lack of accurate reddening estimates (precluded by the
presence of emission lines in both bands) we are unable to quantify this
further.

\begin{deluxetable}{rcc}
\tablewidth{0pt}
\tablecaption{\label{tab:blue_extended}
	Star Formation Parameters for the Extended Blue Continuum}
\tablehead{
Model\phm{mmmmmm}		& $\log(\tau)$ (yr)	&  SFR (\Msun/yr)
}
\startdata
Const. SFR: Salpeter: \Zsun\phm{0.02}	& $8.8 - 9.7$	& $1.1 - 0.8$	\nl
		0.2\Zsun\phm{0}		& $8.7 - 9.7$	& $1.0 - 0.6$	\nl
		0.02\Zsun\phm{}	   & $\phn9.2 - 10.1$	& $0.7 - 0.4$	\nl
	     Scalo: \Zsun\phm{0.02}	& $8.6 - 9.3$	& $2.1 - 1.0$	\nl
		0.2\Zsun\phm{0}		& $8.6 - 9.4$	& $1.7 - 0.7$	\nl
		0.02\Zsun\phm{}		& $8.9 - 9.8$	& $1.0 - 0.4$	\nl
\tableline
					&		& $\log(M/$\Msun)  \nl
\tableline
Inst. Burst: Salpeter: \Zsun\phm{0.02}	& $7.6 - 8.7$	& $8.5 - 9.3$	\nl
		0.2\Zsun\phm{0}		& $7.4 - 8.8$	& $8.3 - 9.3$	\nl
		0.02\Zsun\phm{}		& $8.3 - 9.0$	& $8.8 - 9.1$	\nl
	     Scalo: \Zsun\phm{0.02}	& $7.7 - 8.7$	& $8.7 - 9.3$	\nl
		0.2\Zsun\phm{0}		& $7.5 - 8.9$	& $8.6 - 9.3$	\nl
		0.02\Zsun\phm{}		& $8.4 - 9.1$	& $8.9 - 9.3$
\enddata
\tablecomments{The upper and lower limits of the timescale $\tau$ of each model
correspond to the epochs in the model evolution that produce the observed range
of $m_{\rm F450W} - m_{\rm F702W}$ colors. For the Constant SFR models, the
tabulated SFR is that which is required to match the observed integrated
magnitude of the excess blue continuum. For the Instantaneous Burst models,
we instead tabulate the total mass of stars since in this model all the star
formation occurs at one instant in time. Further details are given in the text.}
\end{deluxetable}

The southern radio lobe displays substantially less blue emission than the
northern lobe. Its $m_{\rm F450W} - m_{\rm F702W}$ color is similar to that of
the northern lobe but its flux is much less, with an integrated apparent
magnitude of only $m_{\rm F450W} \sim 23.2$. Since it is fainter, errors in its
measured properties are much more dominated by uncertainties in the underlying
continuum, reddening correction, and contributions from line emission,
therefore we do not feel justified in carrying out a comparison with star
formation models to the same level of detail as the more prominent northern
blue excess. We do note, however, one interesting feature of the southern blue
emission, namely a blue ``arc'' that follows the outer contours of the southern
lobe, from its western edge around to the south. This is unlikely to correspond
to optical synchrotron emission, as the radio contours show no enhanced
brightening at the location of the arc, thus the arc may also indicate emission
from young stars although our data are of insufficient S/N to quantify this
further.

\subsubsection{The Isolated Blue Knots}

\begin{deluxetable}{lccccc}
\tablewidth{0pt}
\tablecaption{\label{tab:blue_compact}
	Compact Blue Objects in A2597}
\tablehead{
ID	& $\Delta\alpha$	& $\Delta\delta$	&
		$m_{\rm F450W}$	& $m_{\rm F702W}$	&
				$m_{\rm F450W} - m_{\rm F702W}$	\nl
	& (\arcsec)		& (\arcsec)		&
		(mag)		& (mag)			&
				(mag)
}
\startdata
1	& $-1.79$ & $-2.21$ & $26.41 \pm 0.48$	& $25.55 \pm 0.26$	& \phs$0.86 \pm 0.55$	\nl
2	& $-3.39$ & $-1.94$ & $25.38 \pm 0.16$	& $25.07 \pm 0.14$	& \phs$0.31 \pm 0.21$	\nl
3	& $-2.14$ & $-2.99$ & $25.42 \pm 0.18$	& $25.10 \pm 0.16$	& \phs$0.32 \pm 0.24$	\nl
4	& $-3.47$ & $-2.57$ & $25.64 \pm 0.23$	& $25.88 \pm 0.29$	& $-0.24 \pm 0.37$
\enddata
\tablecomments{For each object, the offsets in Right Ascension and Declination
($\Delta\alpha$ and $\Delta\delta$ respectively) are given in arcseconds
relative to the continuum nucleus of the galaxy. The 1~$\sigma$ uncertainties
on the individual magnitudes $m_{\rm F450W}$ and $m_{\rm F702W}$ are derived
from the count-rate statistics, and are combined in quadrature to yield
uncertainties for the color index $m_{\rm F450W} - m_{\rm F702W}$.}
\end{deluxetable}

We now discuss a group of four blue clumps, or knots, that are readily apparent
south-west of the southern radio lobe. Because of their relative isolation,
they can serve as a valuable reference for comparison when discussing the star
formation that has occurred north of the nucleus. In the original (unsmoothed)
PC1 images they are marginally resolved, with FWHM values in the range
$\sim 0\farcs12 - 0\farcs16$. They apparently correspond to the southern
{\it U $-$ I\/} blue continuum excess reported by
	\citeN{McNamara.1993.AJ.105.417}.
These features do not appear to be directly associated with the radio lobe, and
in all likelihood they are young star clusters. In Table~\ref{tab:blue_compact}
we present the magnitudes and colors of these objects, calibrated in the \HST\
VEGAMAG photometric system. Our photometric analysis of these objects indicates
that their colors are about $1.5 - 2$ magnitudes bluer than the surrounding
elliptical continuum. As a consistency check, we also carried out photometry of
a number of the brighter, redder objects around A2597 in the F702W image, and
compared these with the {\it R\/}-band photometry of the same objects presented
by
	\shortciteN{Holtzman.1996.AJ.112.416}.
We found that the magnitudes typically agreed to within $\sim 0.1 - 0.2$~mag,
which is quite satisfactory given the uncertainties in the color terms that had
been used by
	\shortciteN{Holtzman.1996.AJ.112.416}
to transform their \HST\ magnitudes to the Johnson/Cousins system.

We have compared the colors and magnitudes of the blue clumps to the stellar
synthesis models of 
	\citeN{Charlot.1999.prep},
K-corrected to the redshift of A2597 and calibrated in the \HST\ bandpasses as
described in the previous section. Since we are investigating individual
clusters in this case, we carry out a comparison only with the instantaneous
burst models, with the same range of metallicities and IMFs as for the northern
blue continuum. In the lower panel of Figure~\ref{fig:charlot_models} we plot
the evolution in color-magnitude space of three population synthesis models
with a Salpeter IMF, metallicities of \Zsun, 0.2\Zsun\ and 0.02\Zsun, and total
mass $10^5$~\Msun, together with the observed colors and magnitudes of the blue
clusters. The comparison suggests that the colors of the three bluest clusters
are consistent with ages $\gtrsim 5 - 10$~Myr and metallicities above
$\sim 0.2Z_{\sun}$, with ages of the two redder of these three clusters ranging
up to a maximum of $\sim 100$~Myr for masses up to $\sim 5 \times 10^5$~\Msun.
The reddest of the four clusters is either much older ($\sim 10^9$~yr) with a
mass $\sim 10^6$~\Msun, or otherwise is the same age as the other three, but
has higher extinction and/or a metallicity significantly above solar (however,
this cluster is fainter than the other three and its properties are not as well
constrained). In \S~\ref{sec:implications} we discuss in further detail the
implications of the measured properties of the blue continuum regions for
various physical mechanisms that can be responsible for triggering the star
formation.

\subsection{Elliptical Isophote Fits%
\label{sec:ell_fits}}

In order to present a more detailed discussion of the various components that
make up the central galaxy, including the filaments, the blue continuum, and
possible dust obscuration, it is necessary to model and remove the underlying
elliptical stellar continuum distribution. This was done by means of the
IRAF/STSDAS task ELLIPSE (based on an algorithm described by
	\citeNP{Jedrzejewski.1987.MNRAS.226.747}),
which makes use of iterative sigma-clipping algorithms to reject the possible
contribution of non-elliptical components to the fit.

Our data are complicated both by the presence of filaments in emission, and
also by apparent dust absorption in the central regions of the galaxy. We
initially attempted techniques similar to those described by
	\shortciteN{Carollo.1997.ApJ.481.710}
and
	\citeN{Forbes.1995.AJ.109.1988},
involving a number of repeated fits to both the F702W and F450W images, which
allows the possibility of correcting for patches of dust obscuration. However,
this technique does not easily allow additional corrections for emission
features, particularly the line emission from the filaments which is present
in both the F450W and the F702W broad-band images of A2597.

Therefore, we obtained the underlying elliptical continuum by fitting isophotes
to a near-infrared H-band image of A2597, recently obtained with the F160W
filter on NICMOS by
	\shortciteN{Donahue.1998.prep}.
This band has the advantage of containing no line emission from the filaments,
and also with the effect of dust obscuration being much reduced (although still
present). We fitted elliptical isophotes to this image by allowing the flux,
ellipticity and position angle of each isophote to vary freely, and using
sigma-clipping to reject features deviating by more than 3~$\sigma$ from the
mean level in each isophote (for example corresponding to stars or clusters, or
the dust obscuration region). Inspecting the resulting residual image showed
that this model provided a good fit to the overall galaxy profile, with the
only significant residuals corresponding to localized features such as clusters
or dust absorption patches. Some low-level residual structure on large scales
was also present, which is attributed to uncertainties in the flat-fielding
applied in the reduction process.

The elliptical model obtained from the NICMOS data was then applied to the
F702W and F450W images, by keeping the isophote shape parameters constant (\ie
ellipticity and position angle), but allowing the flux of each isophote to vary
freely, and using sigma-clipping as before to remove contributions due to the
filaments or dust patches. For isophotes with a semi-major axis $\gtrsim 6''$
we had to allow the isophote shape parameters to vary, since the smaller size
of the NICMOS image, together with its low-level flat-field effects, produced
inaccuracies in the model at these large radii. However, for isophotes inside
this radius the NICMOS isophote shapes provided a very good fit to the F702W
and F450W data. Subtracting the resulting model from the input image yielded
for each filter a first-iteration residual image which contained most (but not
yet all) of the ``excess'' emission. Because of the substantial contributions in
emission from the filaments within the central $\sim 1 - 2''$, we carried out a
second iteration, according to the following procedure. Pixels deviating from 0
by more than 3 times the rms value in the residual image were subtracted from
the original input image to yield a second input image. New isophotes were then
fitted to this image, using the isophote shape parameters from the first
iteration and allowing only the flux to vary, and again using sigma-clipping to
reject non-elliptical emission features. This yielded a second residual image;
repeating this procedure further showed that the process had converged by the
second iteration to provide an accurate representation of the underlying
elliptical continuum flux distribution, together with the residual emission and
absorption features.

\begin{figure}
\plotfiddle{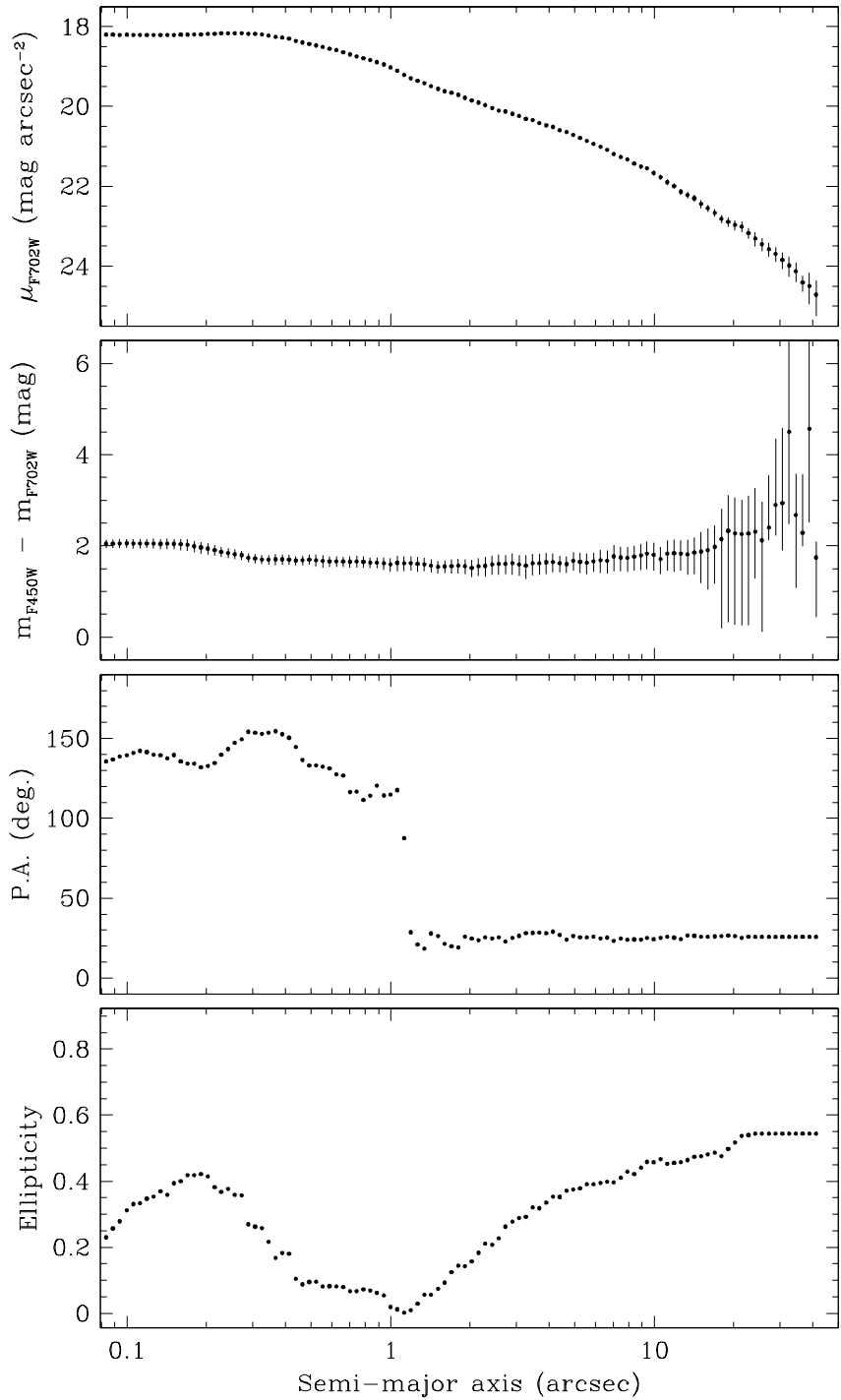}{515pt}{0}{95}{95}{-300pt}{-100pt}
\figcaption[fig5.ps]
{The final set of isophotal parameters fitted to the galaxy continuum
distribution, together with the $m_{\rm F450W} - m_{\rm F702W}$ color
distributions. Error bars are absent from the position angle and ellipticity
plots since these parameters were kept constant in producing the final fits to
the F702W and F450W images.%
\label{fig:isophotes}}
\end{figure}

In Figure~\ref{fig:isophotes} we present the final set of isophotal parameters
fitted to the galaxy. The elliptical continuum color index across most of the
galaxy is in the range $m_{\rm F450W} - m_{\rm F702W} \sim 1.5 - 2$ which
corresponds to {\it B $-$ V\/} $\sim 1 - 1.5$, a typical value for gE galaxies
	(\eg \shortciteNP{Carollo.1997.ApJ.481.710}).
The color is somewhat redder in the central region, possibly related to the
presence of dust. One interesting feature of the galaxy is a change of
$\sim 90\arcdeg$ in the orientation of the major axis, relative to the outer
isophotes, for semi-major axis lengths $\lesssim 1\arcsec$. This also
corresponds to an increase in ellipticity toward the center. This change may be
due to the presence of dust located on either side of the nucleus, although in
the F160W image the change is quite smooth and therefore we cannot rule out the
possibility that this may reflect a genuine shift in the underlying light
distribution, for example due to a nuclear bar
	(\eg \shortciteNP{Nieto.1992.AA.257.97}).

\begin{figure}
\plotone{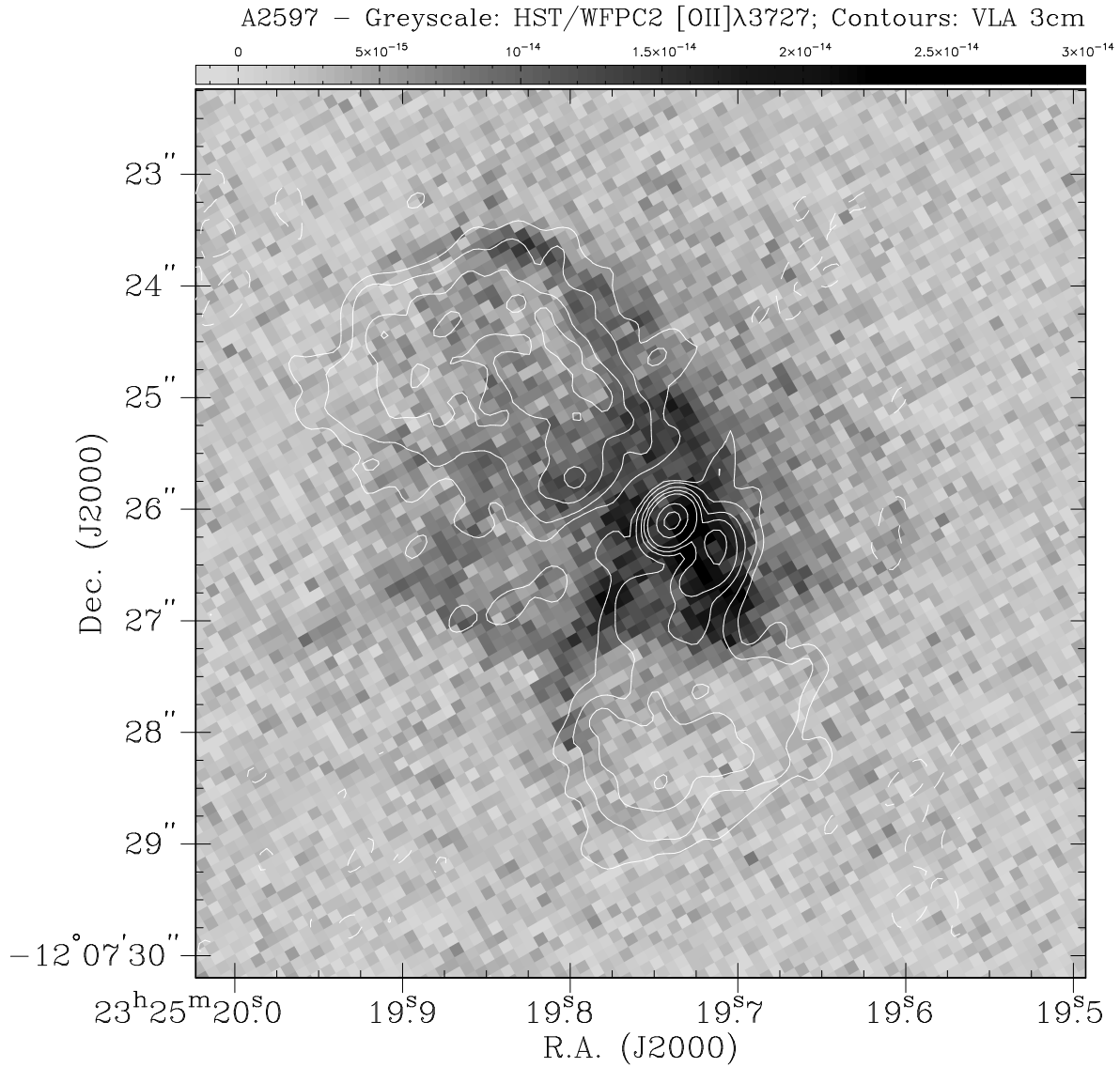}
\figcaption[fig2.ps]
{Narrow-band F410M \HST\ WFPC2 WF3 image of A2597 (greyscale). The spatial
scale is $\sim 0\farcs0996$/pixel, corresponding to $\sim 140$~pc~/ pixel for
our adopted cosmological parameters. At the redshift of
the object, this filter fully samples the [\OII] $\lambda$3727 emission line,
and this image can be considered representative of the emission-line structure,
although we discuss in the text the possibility that continuum emission also
contributes to some of the filament structure. We superpose contours from a
3.5~cm VLA radio continuum image obtained by
	\protect\shortciteN{Sarazin.1995.ApJ.447.559}.
Note the extremely good morphological agreement between the northern radio lobe
and emission-line arc, and also between the southern radio hotspot and the
bright, compact line-emitting knot. The southern emission-line filament also
lies near the eastern edge of the southern radio lobe.%
\label{fig:f450W_3cm}}
\end{figure}

In Figure~\ref{fig:f702W_res} we present the ``residual'' filamentary structure
in the F702W band, produced by subtracting the elliptical galaxy model from the
original image. We also overlay contours representing the VLA 3.5~cm radio
continuum image
	\shortcite{Sarazin.1995.ApJ.447.559}.
The positive emission features in this image are likely to be dominated by \Ha\
line emission rather than stellar continuum, based on the fact that the
filamentary structures in the F410M image (Figure~\ref{fig:f450W_3cm}) are
dominated by [\OII] $\lambda$3727 emission. The filaments in
Figure~\ref{fig:f702W_res} are sampled with both higher S/N and better spatial
resolution than those in Figure~\ref{fig:f450W_3cm}: the bright arcs are
resolved into clumps in several locations, and diffuse emission can be clearly
distinguished across the face of the northern radio lobe. In addition, the
presence of strong, patchy dust obscuration is evident in a number of locations,
particularly to the south-east and to the north of the nucleus. Some weaker
obscuration is also present west of the nucleus, just beyond the bright radio
hotspot and associated emission-line filaments.

\subsection{The Emission-Line Morphology}

\begin{deluxetable}{lcccc}
\tablewidth{0pt}
\tablecaption{Emission-Line Fluxes and Luminosities%
	\label{tab:N_lobe_fluxes}}
\tablehead{
Region		& $F_{\rm[O\;II]\;\lambda3727}$	& $L_{\rm[O\;II]\;\lambda3727}$	& $L_{\rm H\beta}$	\nl
		& (\ergss~cm$^{-2}$)		& (\ergss)			& (\ergss)
}
\startdata
N. arc - upper	& $9.31 \times 10^{-15}$	& $1.30 \times 10^{41}$		& $2.96 \times 10^{40}$	\nl
N. arc - lower	& $9.49 \times 10^{-15}$	& $1.33 \times 10^{41}$		& $3.02 \times 10^{40}$	\nl
N. lobe face	& $1.58 \times 10^{-14}$	& $2.21 \times 10^{41}$		& $5.02 \times 10^{40}$	\nl
N. lobe (total)	& $3.46 \times 10^{-14}$	& $4.84 \times 10^{41}$		& $1.10 \times 10^{41}$	\nl
SW knot		& $9.52 \times 10^{-15}$	& $1.33 \times 10^{41}$		& $3.03 \times 10^{40}$	\nl
SE arc		& $8.45 \times 10^{-15}$	& $1.18 \times 10^{41}$		& $2.68 \times 10^{40}$
\enddata
\tablecomments{The [\OII] $\lambda$3727 line fluxes were obtained by summing
all the pixels in the corresponding regions on the WF3 F410M image, which is
dominated by [\OII] $\lambda$3727 line emission.}
\end{deluxetable}

The distribution of optical line emission around the northern lobe of the radio
source, as displayed in Figures~\ref{fig:f450W_3cm} and \ref{fig:f702W_res},
can be divided into three regions: (1)~the bright ``upper'' arc along the
north-west of the lobe; (2)~a less distinct ``lower'' arc around the southern
boundary of the lobe; (3)~diffuse, lower surface-brightness emission across the
face of the radio lobe. The upper arc runs parallel to a bright ridge of
enhanced radio emission along the edge of the lobe, both of which display a
linear extent of $\sim 2''$ (2.8~kpc). The arcs are dominated by emission-line
radiation; the lack of any detected blue continuum emission excludes
non-thermal processes such as optical synchrotron emission or inverse Compton
scattering. We discuss these features in more detail in \S\ref{sec:N_lobe}.

The southern lobe displays a compact knot of high surface-brightness line
emission co-incident with the strong radio hotspot immediately south-west from
the nucleus. A prominent semi-circular filament of emission is also present SE
from the nucleus, along the eastern edge of the southern radio lobe. We find
that these features are also dominated by line emission and not continuum
radiation. We discuss the southern emission-line features further in
\S\ref{sec:S_knot}. 

In Table~\ref{tab:N_lobe_fluxes} we present the integrated [\OII] $\lambda$3727
fluxes of each of the three northern emission-line regions, together with the
total luminosity of emission-line gas surrounding the northern radio lobe, as
well as the  bright emission-line knot immediately SW from the nucleus and the
semi-circular SE arc. We also derive the \Hb\ luminosities for each of these
regions from the measured [\OII] $\lambda$3727 line fluxes by adopting a
canonical value of \Hb\ / [\OII] $\lambda$3727 = 0.23 (uncorrected for
reddening), determined from the ground-based spectra of
	\citeN{Voit.1997.ApJ.486.242}.
Implicit in this conversion is the assumption that the excitation and
extinction properties of the nebula as a whole are also representative of the
material associated with each radio lobe; this approximation is necessitated by
the current absence of spectra with higher spatial resolution.

\section{Interaction between the Northern Radio Lobe and Ambient Gas%
\label{sec:N_lobe}}

A prominent feature of this galaxy is the complex network of emission-line
filaments associated with the radio lobes. In this section we examine scenarios
in which the northern filaments trace physical interactions between the
northern radio lobe and ambient gas. We derive physical parameters of the
ionized and neutral gas around the radio lobe, including the neutral fraction
and volume filling factor of the gas and its total mass, by combining
constraints obtained from our new emission-line data with those from previous
observations.

We consider a geometry in which the radio lobe is surrounded by an
approximately uniform shell of line-emitting gas (except along the edge
furthest away from the nucleus, where no significant emission is detected).
This scenario is suggested by the similarities in flux and spatial extent of
the upper and lower edge-brightened arcs. Furthermore, the integrated flux of
the diffuse emission across the face of the radio lobe (which would include
emission both from behind and in front of the lobe) is approximately twice that
of the flux in each arc, supporting an edge-brightened geometry for the shell.

Limits on the shell thickness or depth $d$, obtained from the width of the
arcs, are thus applicable to the line-emitting gas covering the face of
the radio lobe. The width of the bright upper arc is only marginally resolved,
covering less than $\sim 2$ PC1 pixels (130~pc), therefore we retain $d$ as a
free parameter normalized to a fiducial value of 100~pc. The surface area of
diffuse line emission across the face of the radio lobe is
$\sim 3.24$~arcsec$^2$, thus the volume of the emission-line region in front of
the radio lobe is $V_{\rm tot} \sim 0.65\, (d / {\rm 100~pc})$ kpc$^3$.
Given standard assumptions such as Case~B ionization conditions, uniform
density emission-line gas, and an edge-brightened shell geometry, our
observed \Hb\ luminosity then provides a combined constraint on the
volume filling factor $f_{\rm H^+}$ and electron density $n_e$ of the warm
ionized gas (with electron temperature $T_e$):
\begin{equation}
\label{eqn:nf_1}
	n_e^2 \, f_{\rm H^+} \sim 11
		\left(  \frac {T_e} {\rm 10^4~K} \right)^{0.87}
		\left(  \frac {d} {\rm 100~pc}  \right)^{-1}
				\hspace{1em} {\rm cm^{-6}}	\, ,
\end{equation}
We aim to derive the physical properties of the gas by combining this
constraint with those obtained from \HI\ observations.
	\shortciteN{ODea.1994.ApJ.436.669}
observed extended, broad \HI\ absorption (${\rm FWHM} \sim 412$~\kms) against
the northern radio lobe. The kinematic properties are similar to those of the
optical line-emitting gas, thus the northern radio lobe would be surrounded by a
collection of clouds containing both neutral and ionized gas. In this section
we derive physical properties of the gas for the following two plausible
scenarios:
\begin{list}
{\arabic{tmpii}.}{\usecounter{tmpii}\setlength{\leftmargin}{0in}
\setlength{\tmplngth}{\parindent}\addtolength{\tmplngth}{\labelwidth}
\addtolength{\tmplngth}{-\labelsep}\setlength{\itemindent}{\tmplngth}}
\item
The \HI\ is co-spatial with the ionized gas, with each \HI\ cloud possessing an
ionized ``skin'' responsible for the line emission
	(\eg \shortciteNP{ODea.1994.ApJ.436.669}).
In this case the ionized and neutral gas are swept up by the expanding radio
lobe into a shell, thereby compressing the gas and determining its kinematics,
but not necessarily ionizing it (alternative ionization mechanisms are
discussed by
	\citeNP{Voit.1997.ApJ.486.242}).
For simplicity, in this scenario we consider the line-emitting gas around the
northern lobe to be entirely photoionized by radiation originating from an
external source (\eg the active nucleus, hot stars, or radiation escaping from
shocks in another region).
\item
The ionized and neutral gas are located in different regions, for example if
the northern radio lobe expands into a previously existing ensemble of neutral
clouds (tidal debris or material condensed out of the cooling flow), fully
ionizing each cloud by means of fast shocks driven into the gas. In this case,
the radio lobe would be surrounded by an inner region of mostly ionized gas and
an outer region of neutral clouds (see for example the schematic presented in
	\citeNP{Bicknell.1997.ApJ.485.112}).
The ionized gas would contain a mixture of cooling ``post-shock'' gas together
with photoionized ``precursor'' gas that absorbs ionizing radiation emitted by
the shocked gas.
\end{list}

\subsection{Scenarios for the Distribution of Neutral and Ionized Gas}

\subsubsection{Co-Spatial Neutral and Ionized Gas}

In this scenario, the neutral and warm ionized components occupy the same
region but are represented by different volume filling factors. Parameterizing
the depth of the \HI\ absorbing material by $d$ thus converts the observed \HI\
column density
($N_{\rm H} \sim 4.5 \times 10^{20} \,(T_s/{\rm 100~K})$~cm$^{-2}$,
	\shortciteNP{ODea.1994.ApJ.436.669})
to a combined constraint on the \HI\ density and volume filling factor:
\begin{equation}
\label{eqn:nf_2}
	n_{\rm H^0} \, f_{\rm H^0} \sim 1.5
		\left(  \frac {T_s} {\rm 100~K}  \right)
		\left(  \frac {d} {\rm 100~pc}  \right)^{-1}
					\hspace{1em} {\rm cm^{-3}}	\, .
\end{equation}
Pressure balance is considered to hold between the neutral gas (spin temperature
$T_s \sim 100$~K), warm ionized gas ($T_e \sim 10^4$~K) and the hot ambient
medium ($T \sim 10^7$~K). We define $\xi = M_{\rm H^0} / M_{\rm H^+}$ to
represent the relative masses of neutral and warm ionized gas, and combine this
with equations (\ref{eqn:nf_1}) and (\ref{eqn:nf_2}) to obtain the gas
densities:
\begin{eqnarray}
	n_{\rm H^+} & \sim & 7.3 \; \xi
		\left(  \frac {T_s} {\rm 100~K}  \right)^{-1}
		\left(  \frac {T_e} {\rm 10^4~K}  \right)^{0.87}
				\hspace{1em} {\rm cm^{-3}}	\, ,
								\nonumber  \\
	n_{\rm H^0} & \sim & 1.5 \times 10^3 \; \xi
		\left(  \frac {T_s} {\rm 100~K}  \right)^{-2}
		\left(  \frac {T_e} {\rm 10^4~K}  \right)^{1.87}
				\hspace{1em} {\rm cm^{-3}}	\, .
\end{eqnarray}
Taking $n_{\rm H^+} \approx n_e \sim 200$~cm$^{-3}$
	\cite{Voit.1997.ApJ.486.242}
to represent the overall line-emitting gas, we obtain the volume filling factor
and thus the total mass of neutral and ionized gas in the shell, assuming only
that its thickness $d$ remains approximately uniform:
\begin{eqnarray}
	M_{\rm H^+} & = & 3.7 \times 10^6 \,
		\left(  \frac {T_e} {\rm 10^4~K}  \right)^{0.87}
		\left(  \frac {n_e} {\rm 200~cm^{-3}}  \right)^{-1}
				\hspace{1em} M_{\sun}	\, ,
								\nonumber  \\
	M_{\rm H^0} & = & 1.0 \times 10^8 \,
		\left(  \frac {T_s} {\rm 100~K}  \right)
				\hspace{1em} M_{\sun}	\, .
\end{eqnarray}

If the gas is distributed in the form of approximately spherical neutral clouds
with ionized skins, then the above constraints yield the relative extent of the
${\rm H^+}$ and ${\rm H^0}$ cloud radius:
\begin{equation}
	\frac {r_{\rm H^+}} {r_{\rm H^0}} =
	\left[1 + 7.3 \times
		\left(  \frac {T_s} {\rm 100~K}  \right)^{-2}
		\left(  \frac {T_e} {\rm 10^4~K}  \right)^{1.87}
		\left(  \frac {n_e} {\rm 200~cm^{-3}}  \right)^{-1}
					\right]^{\frac{1}{3}}	\, .
\end{equation}
We use this, together with the projected surface area of the northern lobe and
the \HI\ volume filling factor, to constrain the \HI\ cloud covering fraction
$c_f$ and cloud radius:
\begin{equation}
	r_{\rm H^0} \, c_f = 2.7 \times 10^{-3}
		\left(  \frac {T_s} {\rm 100~K}  \right)^{2}
		\left(  \frac {T_e} {\rm 10^4~K}  \right)^{-1}
		\left(  \frac {n_e} {\rm 200~cm^{-3}}  \right)^{-1}
				\, {\rm pc}	\, .
\end{equation}
Combining this with the limits on the covering fraction of the extended \HI\
absorption in front of the northern radio lobe ($c_f > 0.006	\, , $
	\shortciteNP{ODea.1994.ApJ.436.669})
and using our determination of the surface area of line-emitting gas in front
of the radio lobe, yields an upper limit on the cloud radius:
$
	r_{\rm H^0} < 0.46 \, {\rm ~pc}	\, ,
$
and corresponding numbers of clouds:
$
	N_{\rm cl} > 6.0 \times 10^4	\, .
$
A covering fraction of unity is obtained with
$N_{\rm cl} \gtrsim 2.8 \times 10^{11}$ clouds, with radii
$r_{\rm H^0} \lesssim 2.7 \times 10^{-3}$~pc.

We conclude our description of the physical properties of the gas in this
scenario with a calculation of the  inferred pressure associated with the gas:
$ P \approx 5.5 \times 10^{-10} \, (n_e / {\rm 200~cm^{-3}}) \, (T_e / {\rm 10^4~K})$ dyn~cm$^{-2}$	\, .
This is in very good agreement with the central X-ray pressure
$P_X \sim 6 \times 10^{-10}$~dyn~cm$^{-2}$ derived by
	\shortciteN{Sarazin.1995.ApJ.447.559}
but is lower than the estimated minimum pressure of the northern lobe,
$P_{\rm me} \sim 1.4 \times 10^{-9}$~dyn~cm$^{-2}$ (re-calculated with the
cosmological parameters used in the present paper), thereby suggesting that
the radio lobe is somewhat overpressured with respect to the surrounding gas.

\subsubsection{Spatially Distinct Neutral and Ionized Components}

An alternative scenario for the distribution of neutral and ionized gas
involves expansion of the radio lobe into a more extended region of neutral
clouds. Each cloud is fully ionized by the passage of the bowshock, analagous
to the scenario envisioned for GPS/CSS radio sources 
	\shortcite{Bicknell.1997.ApJ.485.112}.
The shell of warm ionized gas around the radio lobe would be surrounded by
neutral clouds whose origin may be unassociated with the radio lobe (for
example, tidal debris around the host galaxy, or gas condensed from the cluster
cooling flow). Line emission is produced by cooling ``post-shock'' gas
downstream from the shocks as well as ``precursor'' material immediately
upstream from the
shocks, photoionized by soft X-ray / UV radiation from the post-shock gas. The
ionized gas clouds possess a bi-modal density distribution: the post-shock
clouds have densities $\sim 2$ orders of magnitude above those of the
photoionized precursor gas, while the relative total line radiation output of
the two regions are approximately equal
	\cite{Dopita.1995.ApJ.455.468}.

We consider a case in which the expansion velocity of the emission-line shell
surrounding the radio lobe is equated with a shock velocity
$V_{\rm sh} \approx 300$~\kms\ (\ie half of the observed optical linewidth).
A relatively uniform expansion speed is suggested by the well-ordered
morphology of the emission-line gas. MAPPINGS~II models show that the
post-shock material in this case cools to gas with a density $\sim 230$ times
greater than the precursor material
	(\citeNP{Dopita.1996.ApJS.102.161};
	\citeNP{Koekemoer.1996.PhD};
	\citeNP{Koekemoer.1998.ApJ.497.662}).
The electron density measured from the \SII\ ratio by 
	\citeN{Voit.1997.ApJ.486.242}
represents an average over all the emission from both regions, thus we use
their value together with the ratio of post-shock / precursor line emission to
infer a precursor density $\approx 2$~cm$^{-3}$. Using our observed \Hb\
luminosity, together with the relationships for shock and precursor emission,
yields a required shock surface area of $\sim 25$~kpc$^2$ to account for the
observed line emission. This compares well with the apparent surface area of
the shell ($\sim 30$~kpc$^2$) --- \ie the covering fraction of shocks across
the emission-line shell would be of order unity for precursor gas of this
density. The corresponding spatial scale on which the precursor gas is ionized
would be of order $\lesssim 10 - 100$~pc (depending on its volume filling
factor); beyond this distance the gas would remain largely neutral.

\subsection{A Comparison of the Energy Budget for Each Scenario}

Having calculated the expected physical properties of the gas in each of the
two preceding scenarios, we now compare their relative likelihood. In
particular, we examine whether the radio lobe can supply the required energy to
the surrounding gas.

We begin with the total minimum energy content of the northern radio lobe,
$E_{\rm me} = 1.95 \times 10^{57}$~ergs, the lower limit on the radio source
age $t_{\rm rad} \gtrsim 5 \times 10^6$~yr from the synchrotron loss timescale,
and the radio luminosity of the northern lobe $\sim 4 \times 10^{42}$~\ergss\
(derived from the observed radio properties by
	\shortciteNP{Sarazin.1995.ApJ.447.559},
and rescaled to the cosmological parameters used in this paper). We assume a
fiducial value of $\lesssim 10\%$ for the conversion efficiency of the total
jet kinetic power $L_j$ into radio emission, noting that this value is not
well constrained observationally
	(\citeNP{Eilek.1982.ApJ.254.472};
	\citeNP{Begelman.1984.RevModPh.56.255}),
\ie $L_j \gtrsim 4 \times 10^{43}$~\ergss.

In the first scenario, the energy transfer from the expanding radio cocoon to
the gas is predominantly kinetic. Given the cospatial nature of the gas
components in this scenario, we equate the expansion speed of the optical
line-emitting gas (300~\kms) with that of the \HI. The kinetic energy of the
neutral and warm ionized gas components is then:
\begin{eqnarray}
	KE_{\rm H^+} & = & 3.3 \times 10^{54} \,
		\left(  \frac {n_e} {\rm 200~cm^{-3}}  \right)^{-1}
		\left(  \frac {T_e} {\rm 10^4~K} \right)^{0.87}
		\left(  \frac {V_{\rm FWHM}} {\rm 600~km~s^{-1}}  \right)^2
				\hspace{1em} {\rm ergs}		\, ,
								\nonumber  \\
	KE_{\rm H^0} & = & 9.2 \times 10^{55} \,
		\left(  \frac {T_s} {\rm 100~K}  \right)
		\left(  \frac {V_{\rm FWHM}} {\rm 600~km~s^{-1}}  \right)^2
				\hspace{1em} {\rm ergs}		\, ,
\end{eqnarray}
which is substantially less than the total minimum energy content of the
northern radio lobe. In other words, the rate of energy input required to
supply this amount of kinetic energy to the combined neutral and warm ionized
gas components need only be $\lesssim 5 - 10\%$ of the total energy input into
the radio lobe, therefore it is feasible that the expansion of the radio lobe
can account for the gas kinematics of the northern emission-line filament
system.

For the second scenario, we investigate whether the energy input from the lobe
expansion can feasibly account for the observed line luminosity through
ionization by radiative shocks. If the rate of work done on the ambient medium
by the expansion of the lobe corresponds directly to the total luminosity of
the fully radiative shocks that are produced, then the resulting \Hb\ line
luminosity can be written in terms of the jet energy flux and shock velocity
	\shortcite{Bicknell.1997.ApJ.485.112}:
\begin{equation}
	L_{\rm H\beta} \sim 0.8 \times 10^{41}
		\left(  \frac {V_{\rm sh}} {\rm 300~km~s^{-1}}  \right)^{-0.59}
		\left(  \frac {L_j} {\rm 4 \times 10^{43}~ergs~s^{-1}}
									\right)
				\hspace{1em} {\rm ergs~s^{-1}}		\, ,
\end{equation}
where we have equated the shock velocity $V_{\rm sh}$ with the 300~\kms\
expansion velocity of the emission-line shell surrounding the radio lobe.
Despite the inherent uncertainties in deriving the radio lobe and jet energetic
parameters, this value of $L_{\rm H\beta}$ is in good agreement with our
measured total integrated line luminosity for the northern lobe, presented in
Table~\ref{tab:N_lobe_fluxes}. We note that this does not necessarily imply
that shocks {\em are} the dominant ionization mechanism in the northern lobe;
instead we are stating that shocks, {\em if present}, correspond to a set of
physically plausible values of parameter space. There is some question whether
the overall properties of the entire EELR around A2597 can be explained in
terms of shocks, given the constraints on the electron temperature and gas
ionization state presented by
	\citeN{Voit.1997.ApJ.486.242}.
However, their results apply primarily to the integrated spectrum of the
emission-line nebula, to which the northern lobe contributes only $\sim 20\%$
of the flux, thus local ionizing mechanisms in this region can readily be
accommodated; higher spatial resolution spectra will prove useful in
further addressing such questions.

Thus, we have shown: (1)~it is energetically feasible that the northern radio
lobe can supply the kinetic energy of the surrounding gas, irrespective of the
ionization mechanism; (2)~if the radio lobe expansion indeed produces
auto-ionizing shocks in the gas, then the required shock luminosity and other
properties of the gas are in good agreement with the observations.

\section{Deflection of the Southern Radio Jet%
\label{sec:S_knot}}

The bright radio hotspot 0$\farcs$5 south-west of the nucleus is a region of
great interest since this location corresponds to a sharp change in the
direction of the radio jet toward the east, which also appears to dramatically
affect the resultant morphology of the entire southern radio lobe.
	\shortciteN{Sarazin.1995.ApJ.447.559}
postulated the presence of a dense deflecting clump of cool gas
($T \lesssim 10^4$~K), but the available (ground-based) optical data were of
insufficient spatial resolution to verify its existence. Our \HST\ images
clearly resolve a knot of high surface-brightness optical line emission
associated with the radio hotspot. This object is therefore a highly suitable
candidate for conducting a study of the effects of a direct jet~/ gas cloud
interaction. Here we use our images to infer the physical properties of the gas
clouds, including their size, mass and density (degree of inhomogeneity).

Two plausible scenarios exist for describing the encounter between the jet and
the gas: either the radio jet propagates into a pre-existing region of cool
dense gas, or an initially straight jet is deflected as a result of an
interaction with one or more gas clouds moving directly into it (for example,
due to their orbital motion around the galaxy). The number of free parameters
in each case prohibits a discussion of which scenario is more likely. Rather,
we examine fundamental physical properties of the gas that would be applicable
to either scenario, using considerations based on energetics and momentum
transfer, and following similar arguments to those previously presented by
	\shortciteN{Eilek.1984.ApJ.278.37}
and
	\citeN{Sutherland.1993.ApJ.414.510}.

We consider a non-relativistic jet with fiducial speed $v_j \sim 10^4$~\kms\
and kinetic power $L_j \sim 4 \times 10^{43}$~\ergss. The acceleration imparted
by the jet to gas approximately at rest is related to the jet momentum flux,
thus the minimum gas mass required to significantly deflect the jet over its
lifetime (equated to the radio source age $t_{\rm rad} \sim 5 \times 10^6$~yr)
is $M_{\rm cl} \gtrsim 6 \times 10^6$~$M_{\sun}$. If the gas cloud volume is
approximately spherical (with a diameter given by its transverse dimension,
$d_{\rm cl} \sim 500$~pc), the required mean density of hydrogen nuclei must
exceed $\langle n \rangle \ga 4$~cm$^{-3}$.

We next examine the heating of the gas by the jet, using $\epsilon_H$ to
represent the fraction of $L_j$ which is converted into heat energy. The
cloud needs to be dense enough to cool at this rate or faster, otherwise it
would be heated to high temperatures and disperse. Writing $n_{\rm rms}$ as the
rms average hydrogen number density of the cloud, and the cooling coefficient
as $\Lambda_{22} = \Lambda (T)\, /\, 10^{-22}$~\ergss~cm$^3$, the balance
between the cooling and heating rates requires
\begin{equation}
\label{eqn:n_rms}
	n_{\rm rms} \ga 14 \,
		\left(  \frac{\epsilon_H}{\Lambda_{22}}  \right)^{1/2} \,
		\left(  \frac{L_j}{\rm 4 \times 10^{43}~ergs~s^{-1}}
							\right)^{1/2} \, 
		\left(  \frac{d_{\rm cl}}{\rm 500~pc}
							\right)^{-3/2}
				\, {\rm cm^{-3}}	\, .
\end{equation}
Comparing this with $\langle n \rangle$ suggests that the cloud either exceeds
the lower limit on its mass and/or is inhomogeneous. Finally, the rms electron
density $n_{\rm e,rms}$ in the cloud that is required to produce the observed
\Hb\ line luminosity is
\begin{equation}
\label{eqn:ne_rms}
	n_{\rm e,rms} \approx 11
		\left(  \frac{d_{\rm cl}}{\rm 500~pc}
								\right)^{-3/2}
		\left(  \frac{T_e}{\rm 10^4~K} \right)^{0.43}
				\, {\rm cm^{-3}}	\, .
\end{equation}
The similarity of the densities in equations~(\ref{eqn:n_rms}) and
(\ref{eqn:ne_rms}) might suggest that the deflecting cloud was composed purely
of warm ionized gas at $10^4$~K. However, in photoionized gas at these
temperatures the cooling coefficient is only
$\Lambda (T) \approx 3 \times 10^{-24}$~\ergss~cm$^3$ and such gas could not
radiate energy deposited by the jet unless the efficiency of deposition was
low, $\epsilon_H \approx 0.03$. For typical conditions in photoionized emission
line gas, the bolometric luminosity $L_{\rm bol}$ is about 25 times that in
\Hb, or $L_{\rm bol} \approx 8 \times 10^{41}$~\ergss, which is much smaller
than $L_j$. Thus, either the efficiency of heating the deflecting cloud is low,
or the cloud contains substantial amounts of atomic and/or molecular gas.
The latter scenario is supported independently by the recent \HST\ NICMOS
detection
	\shortcite{Donahue.1998.prep}
of significant molecular H$_2$ emission co-incident with the knot. Hence, we
suggest that the observed properties of this emission-line knot are consistent
with the presence of clumps of dense gas sufficiently massive to deflect the
jet significantly over the lifetime of the radio source.

\section{Implications for Cooling Flows, cD Galaxies and Radio Sources%
\label{sec:implications}}

We have so far demonstrated that the emission-line nebula in the cD elliptical
of A2597 displays a close association with the radio source, and that several
different properties of the gas are strongly suggestive of direct interactions
between the radio lobes and the ambient medium. We have also shown that the
blue continuum emission originates in a number of resolved locations within the
central few kpc, and does not display a direct spatial correspondence with the
emission-line filaments. We have identified a number of regions of recent or
on-going star formation that can be modeled either as single-burst populations
with ages of several times $10^7 - 10^8$~yr, or as continuous (constant) star
formation with ages about an order of magnitude higher. We have also resolved
an elongated region of strong dust obscuration immediately south-east of the
nucleus, together with evidence of patches of dust in other locations nearby.

In this section we explore the likely physical inter-relationships between
these various components in cooling flow clusters, using A2597 as an example.
Studies to date suggest that the strength of the blue continuum excess and
nebular line emission around cD galaxies in cooling flow clusters increases as
a function of the inferred total cooling rate
	(\citeNP{McNamara.1989.AJ.98.2018};
	\citeNP{Allen.1995.MNRAS.276.947};
	\citeNP{McNamara.1997.GCCF.109} and references therein).
Cooling flow cluster cDs possess abnormally high amounts of molecular gas and
dust, and display more compact radio morphologies, relative to cD galaxies
not associated with cooling flow clusters. Thus, it is crucial to understand
the underlying physical connection between the dense, high-pressure ICM
surrounding the elliptical, the blue continuum excess, the nebular line
strength, the strong dust obscuration, and the radio source. In particular, is
the presence of a cooling flow either a necessary and~/ or sufficient
condition to explain these properties of the central galaxy, or do other
processes also play a role (\eg mergers or interactions)? We address this issue
by breaking it down into the following questions: (1)~what has triggered the
recent star formation; (2)~what is responsible for triggering the AGN;
(3)~how are the triggering mechanisms related to the fuel supply in each case,
and what is the relationship between the star formation mechanism, the
triggering of the radio source, and the presence of a cooling flow?

\subsection{What has Triggered the Star Formation?}

Generally, star formation in dense molecular clouds is triggered by compression
of the clouds resulting from enhanced external pressure, or by deep molecular
shocks associated with cloud-cloud collisions. Here we examine the likelihood
of various means of inducing these effects to result in the formation of stars.

One triggering mechanism for star formation is direct interactions between the
radio jets or lobes and dense gas clouds immediately surrounding them
	(\citeNP{Rees.1989.MNRAS.239.1P};
	\citeNP{Begelman.1989.ApJ.345.L21};
	\citeNP{DeYoung.1989.ApJ.342.L59},
	\citeyearNP{DeYoung.1995.ApJ.446.521};
	\citeNP{Daly.1990.ApJ.355.416}).
This scenario invokes the collapse of marginally unstable clouds in the ambient
medium, resulting from overpressurization as they are overtaken by the
expanding radio cocoon. It predicts that, while star formation is initiated
along the radio lobe boundaries, the spatial correlation is short-lived due to
differences between the orbits of the stars and the motion of the radio plasma.
The relatively mild expansion velocity of the cocoon, compared to the working
surface of the jet, prevents disruption of the clouds and facilitates their
collapse. The large-scale asymmetry of the blue continuum in A2597 would
correspond to an intrinsic asymmetry in the initial distribution of molecular
gas clouds. The radio source age in this scenario would have to agree with the
ages of the young stars, which we find to be in the range $\sim 10^7 - 10^8$~yr
from the instantaneous burst stellar model synthesis ages. The lower limit on
the radio source lifetime is $\gtrsim 5 \times 10^6$~yr from the synchrotron
spectral ageing analysis, but given the uncertainties involved in this
derivation the actual radio age can be substantially higher. Thus, from
timescale arguments it is plausible that the star formation is triggered by the
radio source. This would imply that the radio source is intrinsically old but
unable to penetrate the dense surrounding ICM to scales of more than a few kpc.

Star formation in cooling flows is also proposed to occur via pressure-driven
collapse of filaments or clouds condensing out of the cooling flow
	(\shortciteNP{Fabian.1986.ApJ.305.9};
	\citeNP{McNamara.1989.AJ.98.2018};
	\citeNP{Allen.1995.MNRAS.276.947}).
The inferred cooling flow mass accretion rates for A2597
	(\shortciteNP{Crawford.1989.MNRAS.236.277};
	\shortciteNP{Edge.1992.MNRAS.258.177})
are about two orders of magnitude above our values of the SFR. Such
discrepancies are well-known from previous studies of A2597 and other
cooling-flow clusters. Possible solutions include erroneous estimates of the
SFR, for example due to a time-variable SFR or changing IMFs; or mass
deposition into material that does not necessarily form stars. These and other
possibilities are discussed in detail by
	\citeN{Johnstone.1987.MNRAS.224.75};
	\citeN{OConnell.1989.AJ.98.180};
	\citeN{Thomas.1990.MNRAS.246.156};
and	\citeN{McNamara.1994.AA.281.673}.
We do not rule this out as a potential source of fuel for star formation.
It does not readily predict, however, the highly asymmetric structure of the
blue continuum in A2597. One means of obtaining asymmetry is via subcluster
merging activity
	(\eg \citeNP{Crawford.1992.MNRAS.259.265},
	\citeyearNP{Crawford.1993.MNRAS.265.431};
	\citeNP{Daines.1994.MNRAS.268.1060}).
However, A2597 displays little evidence of a recent cluster/subcluster merger,
as the X-ray data suggest a relaxed system
	\cite{Sarazin.1997.ApJ.480.203}.
The spatial scale of the blue continuum emission is also one to two orders of
magnitude smaller, $r \lesssim 3$~kpc, compared with the typical cooling radius
of $\sim 50 - 150$~kpc at which the stars are expected to form
	\shortcite{Johnstone.1987.MNRAS.224.75}.
Thus, cooling of the X-ray gas can provide a source of cooler material for
star formation, but an additional mechanism appears to be required for
triggering the concentrated massive star formation observed in A2597.

This leads us to discuss star formation triggered via accretion of a gas-rich
companion galaxy by the elliptical. Galaxy interactions are generally
associated with triggering star formation
	(\eg \citeNP{Toomre.1972.ApJ.178.623};
see also
	\citeNP{Balick.1982.ARAA.20.431},
and
	\citeNP{Barnes.1992.ARAA.30.705}
for reviews). Detailed n-body/hydrodynamic models show that interactions
between ellipticals and small gas-rich galaxies generally lead to gas accretion
into the elliptical core
	\cite{Weil.1993.ApJ.405.142},
triggering a centrally condensed starburst as well as star formation in the
tidal debris
	(\eg \citeNP{Mihos.1994.ApJ.425.L13},
	\citeyearNP{Mihos.1994.ApJ.437.611},
	\citeyearNP{Mihos.1996.ApJ.464.641}).
Such features are observed in a wide range of galaxies, but specifically
in large ellipticals
	(\eg \citeNP{Strom.1979.AJ.84.1091};
	\citeNP{McNamara.1992.ApJ.393.579};
	\shortciteNP{Holtzman.1996.AJ.112.416};
	\citeNP{Sparks.1997.ApJ.486.253}).
The star-forming regions in A2597 show similar characteristics: intense
star formation near the galaxy core, with smaller, more isolated blue clusters
further out. The primary objection to transporting gas-rich galaxies into the
cluster core is that ram-pressure stripping by the hot ICM removes much of the
relatively diffuse \HI\ gas
	(\citeNP{Gunn.1972.ApJ.176.1};
	\citeNP{Nulsen.1982.MNRAS.198.1007}).
However, dense clouds of molecular gas and dust have much lower filling factors
and higher column densities than the \HI\ component and are less affected by
hydrodynamical processes
	(\citeNP{Kritsuk.1983.Afz.19.471};
	\citeNP{Valluri.1990.ApJ.357.367}).
Indeed, studies of the molecular content of galaxies in clusters show no
significant dependence on location in the cluster, whereas the \HI\ content
decreases substantially toward the cluster core
	(\citeNP{Kenney.1989.ApJ.344.171};
	\shortciteNP{Casoli.1991.AA.249.359};
	\citeNP{Lavezzi.1998.AJ.115.405}).
Therefore such an interaction can quite plausibly provide a fresh supply of
gas, predominantly molecular. The stellar model ages of $\sim 10^8 - 10^9$~yr
are also typical of the dynamical timescales involved in such interactions.

Thus, we propose that two plausible triggering mechanisms for the star
formation are: (1)~related directly to the passage of the radio source, or
(2)~related to accretion of a gas-rich companion galaxy. If the star formation
is modeled in terms of a single burst then the resulting ages are in the range
$\sim 10^7 - 10^8$~yr --- the low end of this range is comparable to the lower
limit on the radio source age, thus we associate the single-burst models with
triggering by the radio source. If, on the other hand, the star formation is
modeled using a constant SFR, then the required ages of the stars are much
higher, of order a few times $10^8 - 10^9$~yr. This is too high to be accounted
for in terms of radio source triggering, but is consistent with the dynamical
timescales typically required for interactions between galaxies. While our
current data do not allow us to distinguish further between these two possible
triggering mechanisms, we are able to conclude that triggering by the radio
source is consistent with a single burst $\sim 10^7 - 10^8$~yr ago, while
triggering by an interaction corresponds to continuous, on-going star formation
over longer timescales up to $\sim 10^9$~yr.

\subsection{AGN Trigger Mechanism, Fuel Supply, and Origin of the Gas
Filaments}

Having discussed the star formation trigger mechanism, we now investigate the
trigger for the AGN itself. The majority of cD ellipticals in cooling flow
clusters are known to host radio-loud sources, compared to a much lower
fraction of cD's in non cooling flow clusters
	\cite{Burns.1990.AJ.99.14}.
A number of studies have discussed cooling flows as the primary explanation for
the observed nuclear activity and star formation, through accretion of the
cooling filaments into the central galaxy
	(\eg \shortciteNP{Cowie.1983.ApJ.272.29};
	\shortciteNP{Canizares.1987.ApJ.312.503},
	\citeNP{Fabian.1990.MNRAS.247.439}).
However, cooling flows need not be the sole mechanism responsible for
triggering and fueling the AGN --- in particular, the long timescales of a
steady cooling flow cannot be reconciled with the comparatively short radio
source lifetimes without invoking additional mechanisms to trigger the onset of
the radio source. Furthermore, many AGN with similar properties are found in
host galaxies not associated with cooling flow clusters.

Thus we are motivated to discuss the plausibility of an additional mechanism
which is generally associated with AGN triggering and fueling, namely
interactions between galaxies
	(\eg \citeNP{Balick.1982.ARAA.20.431};
	\citeNP{Barnes.1992.ARAA.30.705},
and references therein). Specifically, ``minor mergers'' between the host
galaxy and a small gas-rich companion are an important means of dumping fuel
into the central region
	(\citeNP{Hernquist.1995.ApJ.448.41};
	\citeNP{Walker.1996.ApJ.460.121};
	\citeNP{DeRobertis.1998.ApJ.496.93}).
In such cases the host galaxy retains its original morphology more
or less intact, with the primary change consisting of a fresh supply of
$\sim 10^8 - 10^9$~\Msun\ of stars, gas and dust.

The central elliptical in A2597 displays a number of features characteristic
of a recent ``minor merger'': (1)~substantial dust which appears dynamically
unsettled, located predominantly on one side of the nucleus; (2)~the extended
complex of neutral, ionized and molecular gas, showing strong asymmetries on
small scales (the present work) and large scales
	\shortcite{Heckman.1989.ApJ.338.48};
(3)~a radio source with an age $\gtrsim 10^7$~yr; (4)~small galaxies nearby,
which if dynamically associated with the cD would have supplied the gas; (5)~the
clusters of young stars. The high-pressure cluster environment associated with
a cooling flow can enhance the star formation to the observed levels and
confine the radio source. However, the actual {\em triggering} mechanism of
the star formation and AGN is plausibly accounted for by this type of
interaction, likely with a galaxy containing substantial amounts of molecular
gas as discussed in the previous section. When such a galaxy interacts with the
central elliptical, some of its surviving molecular material would be disrupted
and become visible as atomic/ionized gas, while the remainder is accreted into
the core of the cD.

In this scenario one expects comparatively little reddening in the diffuse
line-emitting filaments, due to the relatively short sputtering times of dust
in the high-pressure surrounding hot gas
	\cite{Draine.1979.ApJ.231.77};
any remaining dust is concentrated in regions containing substantial amounts of
dense molecular gas. Indeed, we find that the strongest extinction in A2597 is
concentrated in the compact, dusty regions immediately adjacent to the nucleus,
located in the same vicinity as the molecular H$_2$ emission detected by
	\shortciteN{Donahue.1998.prep}.
It is therefore plausible that the observed distribution and physical
properties of the dust and gas in A2597 can be accounted for by the capture of
an infalling gas-rich galaxy which has retained a significant fraction of its
original dense clouds of molecular gas and dust. This would also provide a
supply of fuel for the current phase of activity of the AGN. The age of the
radio source is somewhat younger than the timescale associated with the
interaction, but this likely represents the additional time required to funnel
gas into the center of the galaxy, which in numerical simulations typically
occurs well into the merger process; alternatively the currently observed
radio source may be only the latest in a series of multiple outbursts during
the galaxy interaction process. Finally, the location of this object in the
center of a dense, high-pressure ICM would account for the enhanced star
formation, as well as the apparent strong confinement of the radio source.

\section{Summary and Final Comments}

In this paper we have presented \HST\ WFPC2 images that spatially resolve some
of the most intriguing features of the cD elliptical in A2597:
\begin{list}
{\arabic{tmpii}.}{\usecounter{tmpii}\setlength{\leftmargin}{0in}
\setlength{\tmplngth}{\parindent}\addtolength{\tmplngth}{\labelwidth}
\addtolength{\tmplngth}{-\labelsep}\setlength{\itemindent}{\tmplngth}}
\item
A complex system of emission-line filaments occupies the inner few kpc of the
galaxy, displaying a number of close spatial associations with the
synchrotron-emitting radio lobes. The morphologies of several of the filaments
are strongly suggestive of direct interactions between the radio source and
ambient gas. In particular, we find that the detailed physical properties of
the line-emitting gas around the northern radio lobe can be described in terms
of a shell of neutral and ionized gas that surrounds the expanding radio lobe,
and we show that the energetics of this shell can be readily accounted for by
energy input from the expanding radio lobe. We also resolve a knot of high
surface-brightness line emission approximately co-incident with a radio hotspot
to the south-west of the nucleus, and we show that its properties are
consistent with those of a dense, inhomogeneous clump of gas
($n_{\rm rms} \gtrsim 14$~cm$^{-3}$, $M_{\rm cl} \gtrsim 6 \times 10^6$\Msun)
sufficiently massive to deflect the radio jet over the lifetime of the radio
source and account for the severely disrupted morphology of the southern
radio lobe. Finally, the lack of detectable line emission in a number of
locations around the radio lobes suggests an intrinsically asymmetric
distribution of ambient gas about the galaxy.
\item
The excess blue continuum emission is clearly resolved by our \HST\ images into
a number of components, specifically an intense band of blue emission
associated with the base of the northern radio lobe, as well as blue clusters
south-west from the southern radio lobe, and a number of more diffuse regions
distributed about the radio lobes. The morphological properties of the blue
continuum (together with its low observed polarization,
	\citeNP{McNamara.1998.prep})
rule out several explanations for its origin, including optical synchrotron
or inverse Compton emission, or scattered nuclear light, leaving recent
star-forming regions as the most likely remaining possibility. Comparing our
photometry with stellar population synthesis models shows that these regions
can be modeled either as single-burst populations with ages of several times
$10^7 - 10^8$~yr, or as continuous (constant) star formation with ages about an
order of magnitude higher, depending upon details of the assumed metallicity
and IMF parameters.
\item
The strong dust obscuration in the central region is resolved into a number of
large clumps, each several hundred pc in size. Most of these are distributed in
a region extending away from the nucleus toward the south-east, while another
strong clump of obscuration is evident immediately to the north of the nucleus,
and some less heavily absorbed regions are present just west of the bright
radio~/ emission-line hotspot in the south. A striking feature of the dust is
its strong degree of clumping and highly asymmetric spatial distribution about
the galaxy center.
\end{list}

The cD elliptical in the cooling flow cluster A2597 is a complex system: it
contains a powerful, compact radio source, surrounded by an extensive network
of emission-line filaments with significant amounts of dust and prominent
regions of recent or on-going star formation. We suggest that the high-pressure
environment associated with the cooling flow is crucial to enhancing the star
formation rate to the high levels observed. However, while a steady cooling
flow can also provide a source of fuel for the star formation and the AGN, its
long timescales are difficult to reconcile with the short radio source lifetime,
the substantial amounts of dust and molecular gas in the central few kpc, and
the highly asymmetric distribution of the gas, dust and young stars without
invoking an additional, more recent mechanism to account for these features.

We propose that a recent ``minor merger'' or interaction in which the cD
accreted gas and dust from a smaller gas-rich galaxy provides a plausible
mechanism, specifically in terms of triggering the formation of the radio
source and introducing large amounts of molecular gas and dust into the cD
elliptical. The observed asymmetry of the gas and dust would indicate that the
system is still in the process of virialization, \ie the event took place
within the last $\sim 10^8 - 10^9$~yr. This dynamical timescale is also
compatible with the timescales derived from the stellar models involving
continuous star formation with a constant SFR, \ie a scenario in which the
interaction is responsible for inducing on-going star formation. However, if
the star formation is described in terms of a single burst then the models
imply ages of the order $\sim 10^7 - 10^8$~yr, in which case the star
formation would more likely have been triggered by the passage of the radio
source. The age of the radio source is substantially less than the galaxy
interaction timescale, and this either represents the additional time
required to funnel gas into the center of the cD, or otherwise indicates
that the present phase of radio activity is the latest in a series of
multiple outbursts.

The properties of A2597 are directly tied to the general question of the
relationship between galaxy interactions, nuclear activity and star formation.
If active galaxies in cooling flows such as A2597 are triggered in the same way
as AGN unassociated with cooling flows, then the primary reason for the
difference in their properties, specifically the compact radio morphology and
enhanced star formation, results from being located at the center of a dense,
high-pressure ICM. Better understanding the effect of environment upon star
formation would be especially desirable in the context of studying objects in a
range of environments at low redshifts, as well as investigating the formation
of stars and galaxies at early epochs where the environment is likely to play
an important, perhaps dominant, role.

\acknowledgements
We are grateful to Stephane Charlot and Gustavo Bruzual for kindly making
available their latest stellar synthesis libraries with appropriate
K-corrections and magnitudes calculated in the \HST\ filters that were used to
observe this object. A.~M.~K. would like to thank Stefano Casertano, Brad
Whitmore, and Paul Goudfrooij for a number of valuable discussions, and Andy
Fruchter for helpful explanations concerning use of the DITHER package. We
thank Michael Wise for his involvement in the initial phase of the project.
We also thank the anonymous referee for useful comments which helped to improve
the paper. Partial support for this work was provided by NASA through grant
number \mbox{GO-06717.01-95A} from the Space Telescope Science Institute, which
is operated by AURA, Inc., under NASA contract \mbox{NAS 5-26555}. C.~L.~S. was
supported in part by NASA ASCA grant \mbox{NAG 5-4516} and NASA Astrophysical
Theory Program grant \mbox{5-3057}.

\clearpage

%
%

\begin{thebibliography}{}

\bibitem[\protect\citeauthoryear{{Allen}}{{Allen}}{1995}]{Allen.1995.MNRAS.276%
.947}
{Allen}, S.~W. 1995, MNRAS, 276, 947

\bibitem[\protect\citeauthoryear{{Balick} \& {Heckman}}{{Balick} \&
  {Heckman}}{1982}]{Balick.1982.ARAA.20.431}
{Balick}, B.,  \& {Heckman}, T.~M. 1982, ARA\&A, 20, 431

\bibitem[\protect\citeauthoryear{{Barnes} \& {Hernquist}}{{Barnes} \&
  {Hernquist}}{1992}]{Barnes.1992.ARAA.30.705}
{Barnes}, J.~E.,  \& {Hernquist}, L. 1992, ARA\&A, 30, 705

\bibitem[\protect\citeauthoryear{{Begelman}, {Blandford}, \& {Rees}}{{Begelman}
  et~al.}{1984}]{Begelman.1984.RevModPh.56.255}
{Begelman}, M.~C., {Blandford}, R.~D.,  \& {Rees}, M.~J. 1984, Rev. Mod. Phys.,
  56, 255

\bibitem[\protect\citeauthoryear{{Begelman} \& {Cioffi}}{{Begelman} \&
  {Cioffi}}{1989}]{Begelman.1989.ApJ.345.L21}
{Begelman}, M.~C.,  \& {Cioffi}, D.~F. 1989, ApJ, 345, L21

\bibitem[\protect\citeauthoryear{{Bicknell}, {Dopita}, \& {O'Dea}}{{Bicknell}
  et~al.}{1997}]{Bicknell.1997.ApJ.485.112}
{Bicknell}, G.~V., {Dopita}, M.~A.,  \& {O'Dea}, C.~P. 1997, ApJ, 485, 112

\bibitem[\protect\citeauthoryear{{Biretta et al.}}{{Biretta et
  al.}}{1996}]{Biretta.1996.WFPC2.STScI}
{Biretta et al.} 1996, WFPC2 Instrument Handbook, Version 4.0 (Baltimore:
  STScI)

\bibitem[\protect\citeauthoryear{{Bruzual} \& {Charlot}}{{Bruzual} \&
  {Charlot}}{1993}]{Bruzual.1993.ApJ.405.538}
{Bruzual}, A.~G.,  \& {Charlot}, S. 1993, ApJ, 405, 538

\bibitem[\protect\citeauthoryear{{Burns}}{{Burns}}{1990}]{Burns.1990.AJ.99.14}
{Burns}, J.~O. 1990, AJ, 99, 14

\bibitem[\protect\citeauthoryear{{Canizares}, {Fabbiano}, \&
  {Trinchieri}}{{Canizares} et~al.}{1987}]{Canizares.1987.ApJ.312.503}
{Canizares}, C.~R., {Fabbiano}, G.,  \& {Trinchieri}, G. 1987, ApJ, 312, 503

\bibitem[\protect\citeauthoryear{{Carollo} et~al.}{{Carollo}
  et~al.}{1997}]{Carollo.1997.ApJ.481.710}
{Carollo}, C.~M., {Franx}, M., {Illingworth}, G.~D.,  \& {Forbes}, D.~A. 1997,
  ApJ, 481, 710

\bibitem[\protect\citeauthoryear{{Casoli} et~al.}{{Casoli}
  et~al.}{1991}]{Casoli.1991.AA.249.359}
{Casoli}, F., {Boisse}, P., {Combes}, F.,  \& {Dupraz}, C. 1991, A\&A, 249, 359

\bibitem[\protect\citeauthoryear{{Charlot} \& {Bruzual}}{{Charlot} \&
  {Bruzual}}{1991}]{Charlot.1991.ApJ.367.126}
{Charlot}, S.,  \& {Bruzual}, A.~G. 1991, ApJ, 367, 126

\bibitem[\protect\citeauthoryear{{Charlot} \& {Bruzual}}{{Charlot} \&
  {Bruzual}}{1999}]{Charlot.1999.prep}
{Charlot}, S.,  \& {Bruzual}, A.~G. 1999, in preparation

\bibitem[\protect\citeauthoryear{{Cowie} \& {Binney}}{{Cowie} \&
  {Binney}}{1977}]{Cowie.1977.ApJ.215.723}
{Cowie}, L.~L.,  \& {Binney}, J. 1977, ApJ, 215, 723

\bibitem[\protect\citeauthoryear{{Cowie} et~al.}{{Cowie}
  et~al.}{1983}]{Cowie.1983.ApJ.272.29}
{Cowie}, L.~L., {Hu}, E.~M., {Jenkins}, E.~B.,  \& {York}, D.~G. 1983, ApJ,
  272, 29

\bibitem[\protect\citeauthoryear{{Crawford} \& {Fabian}}{{Crawford} \&
  {Fabian}}{1992}]{Crawford.1992.MNRAS.259.265}
{Crawford}, C.~S.,  \& {Fabian}, A.~C. 1992, MNRAS, 259, 265

\bibitem[\protect\citeauthoryear{{Crawford} \& {Fabian}}{{Crawford} \&
  {Fabian}}{1993}]{Crawford.1993.MNRAS.265.431}
{Crawford}, C.~S.,  \& {Fabian}, A.~C. 1993, MNRAS, 265, 431

\bibitem[\protect\citeauthoryear{{Crawford} et~al.}{{Crawford}
  et~al.}{1989}]{Crawford.1989.MNRAS.236.277}
{Crawford}, C.~S., {Fabian}, A.~C., {Johnstone}, R.~M.,  \& {Arnaud}, K.~A.
  1989, MNRAS, 236, 277

\bibitem[\protect\citeauthoryear{{Daines}, {Fabian}, \& {Thomas}}{{Daines}
  et~al.}{1994}]{Daines.1994.MNRAS.268.1060}
{Daines}, S.~J., {Fabian}, A.~C.,  \& {Thomas}, P.~A. 1994, MNRAS, 268, 1060

\bibitem[\protect\citeauthoryear{{Daly}}{{Daly}}{1990}]{Daly.1990.ApJ.355.416}
{Daly}, R.~A. 1990, ApJ, 355, 416

\bibitem[\protect\citeauthoryear{{De Robertis}, {Yee}, \& {Hayhoe}}{{De
  Robertis} et~al.}{1998}]{DeRobertis.1998.ApJ.496.93}
{De Robertis}, M.~M., {Yee}, H. K.~C.,  \& {Hayhoe}, K. 1998, ApJ, 496, 93

\bibitem[\protect\citeauthoryear{{De Young}}{{De
  Young}}{1989}]{DeYoung.1989.ApJ.342.L59}
{De Young}, D.~S. 1989, ApJ, 342, L59

\bibitem[\protect\citeauthoryear{{De Young}}{{De
  Young}}{1995}]{DeYoung.1995.ApJ.446.521}
{De Young}, D.~S. 1995, ApJ, 446, 521

\bibitem[\protect\citeauthoryear{{Donahue et al.}}{{Donahue et
  al.}}{1998}]{Donahue.1998.prep}
{Donahue et al.} 1998, in preparation

\bibitem[\protect\citeauthoryear{{Dopita} \& {Sutherland}}{{Dopita} \&
  {Sutherland}}{1995}]{Dopita.1995.ApJ.455.468}
{Dopita}, M.~A.,  \& {Sutherland}, R.~S. 1995, ApJ, 455, 468

\bibitem[\protect\citeauthoryear{{Dopita} \& {Sutherland}}{{Dopita} \&
  {Sutherland}}{1996}]{Dopita.1996.ApJS.102.161}
{Dopita}, M.~A.,  \& {Sutherland}, R.~S. 1996, ApJS, 102, 161

\bibitem[\protect\citeauthoryear{{Draine} \& {Salpeter}}{{Draine} \&
  {Salpeter}}{1979}]{Draine.1979.ApJ.231.77}
{Draine}, B.~T.,  \& {Salpeter}, E.~E. 1979, ApJ, 231, 77

\bibitem[\protect\citeauthoryear{{Edge}, {Stewart}, \& {Fabian}}{{Edge}
  et~al.}{1992}]{Edge.1992.MNRAS.258.177}
{Edge}, A.~C., {Stewart}, G.~C.,  \& {Fabian}, A.~C. 1992, MNRAS, 258, 177

\bibitem[\protect\citeauthoryear{Eilek}{Eilek}{1982}]{Eilek.1982.ApJ.254.472}
Eilek, J.~A. 1982, ApJ, 254, 472

\bibitem[\protect\citeauthoryear{{Eilek} et~al.}{{Eilek}
  et~al.}{1984}]{Eilek.1984.ApJ.278.37}
{Eilek}, J.~A., {Burns}, J.~O., {O'Dea}, C.~P.,  \& {Owen}, F.~N. 1984, ApJ,
  278, 37

\bibitem[\protect\citeauthoryear{{Fabian}}{{Fabian}}{1994}]{Fabian.1994.ARAA.3%
2.277}
{Fabian}, A.~C. 1994, ARA\&A, 32, 277

\bibitem[\protect\citeauthoryear{{Fabian} et~al.}{{Fabian}
  et~al.}{1986}]{Fabian.1986.ApJ.305.9}
{Fabian}, A.~C., {Arnaud}, K.~A., {Nulsen}, P. E.~J.,  \& {Mushotzky}, R.~F.
  1986, ApJ, 305, 9

\bibitem[\protect\citeauthoryear{{Fabian} \& {Crawford}}{{Fabian} \&
  {Crawford}}{1990}]{Fabian.1990.MNRAS.247.439}
{Fabian}, A.~C.,  \& {Crawford}, C.~S. 1990, MNRAS, 247, 439

\bibitem[\protect\citeauthoryear{{Fabian} \& {Nulsen}}{{Fabian} \&
  {Nulsen}}{1977}]{Fabian.1977.MNRAS.180.479}
{Fabian}, A.~C.,  \& {Nulsen}, P. E.~J. 1977, MNRAS, 180, 479

\bibitem[\protect\citeauthoryear{{Forbes}, {Franx}, \& {Illingworth}}{{Forbes}
  et~al.}{1995}]{Forbes.1995.AJ.109.1988}
{Forbes}, D.~A., {Franx}, M.,  \& {Illingworth}, G.~D. 1995, AJ, 109, 1988

\bibitem[\protect\citeauthoryear{{Fruchter} \& {Hook}}{{Fruchter} \&
  {Hook}}{1997}]{Fruchter.1997.SPIE.3164.120}
{Fruchter}, A.~S.,  \& {Hook}, R.~N. 1997, in Applications of Digital Image
  Processing XX, Proc. SPIE, Vol. 3164, ed. A.~Tescher (SPIE), 120,
  astro-ph/9708242

\bibitem[\protect\citeauthoryear{{Gunn} \& {Gott}}{{Gunn} \&
  {Gott}}{1972}]{Gunn.1972.ApJ.176.1}
{Gunn}, J.~E.,  \& {Gott}, J.~R. 1972, ApJ, 176, 1

\bibitem[\protect\citeauthoryear{{Heckman}}{{Heckman}}{1981}]{Heckman.1981.ApJ%
.250.L59}
{Heckman}, T.~M. 1981, ApJ, 250, L59

\bibitem[\protect\citeauthoryear{{Heckman} et~al.}{{Heckman}
  et~al.}{1989}]{Heckman.1989.ApJ.338.48}
{Heckman}, T.~M., {Baum}, S.~A., {van Breugel}, W. J.~M.,  \& {McCarthy}, P.
  1989, ApJ, 338, 48

\bibitem[\protect\citeauthoryear{{Hernquist} \& {Mihos}}{{Hernquist} \&
  {Mihos}}{1995}]{Hernquist.1995.ApJ.448.41}
{Hernquist}, L.,  \& {Mihos}, J.~C. 1995, ApJ, 448, 41

\bibitem[\protect\citeauthoryear{{Holtzman} et~al.}{{Holtzman}
  et~al.}{1995a}]{Holtzman.1995.PASP.107.156}
{Holtzman}, J.~A., et~al. 1995a, PASP, 107, 156

\bibitem[\protect\citeauthoryear{{Holtzman} et~al.}{{Holtzman}
  et~al.}{1995b}]{Holtzman.1995.PASP.107.1065}
{Holtzman}, J.~A., {Burrows}, C.~J., {Casertano}, S., {Hester}, J.~J.,
  {Trauger}, J.~T., {Watson}, A.~M.,  \& {Worthey}, G. 1995b, PASP, 107, 1065

\bibitem[\protect\citeauthoryear{{Holtzman} et~al.}{{Holtzman}
  et~al.}{1996}]{Holtzman.1996.AJ.112.416}
{Holtzman}, J.~A., et~al. 1996, AJ, 112, 416

\bibitem[\protect\citeauthoryear{{Hu}, {Cowie}, \& {Wang}}{{Hu}
  et~al.}{1985}]{Hu.1985.ApJS.59.447}
{Hu}, E.~M., {Cowie}, L.~L.,  \& {Wang}, Z. 1985, ApJS, 59, 447

\bibitem[\protect\citeauthoryear{{Jedrzejewski}}{{Jedrzejewski}}{1987}]{Jedrze%
jewski.1987.MNRAS.226.747}
{Jedrzejewski}, R.~I. 1987, MNRAS, 226, 747

\bibitem[\protect\citeauthoryear{{Johnstone}, {Fabian}, \&
  {Nulsen}}{{Johnstone} et~al.}{1987}]{Johnstone.1987.MNRAS.224.75}
{Johnstone}, R.~M., {Fabian}, A.~C.,  \& {Nulsen}, P. E.~J. 1987, MNRAS, 224,
  75

\bibitem[\protect\citeauthoryear{{Kenney} \& {Young}}{{Kenney} \&
  {Young}}{1989}]{Kenney.1989.ApJ.344.171}
{Kenney}, J. D.~P.,  \& {Young}, J.~S. 1989, ApJ, 344, 171

\bibitem[\protect\citeauthoryear{{Koekemoer}}{{Koekemoer}}{1996}]{Koekemoer.19%
96.PhD}
{Koekemoer}, A.~M. 1996, Ph.D. thesis, Australian National University

\bibitem[\protect\citeauthoryear{{Koekemoer} \& {Bicknell}}{{Koekemoer} \&
  {Bicknell}}{1998}]{Koekemoer.1998.ApJ.497.662}
{Koekemoer}, A.~M.,  \& {Bicknell}, G.~V. 1998, ApJ, 497, 662

\bibitem[\protect\citeauthoryear{{Krist} \& {Hook}}{{Krist} \&
  {Hook}}{1997}]{Krist.1997.TinyTim.STScI}
{Krist}, J.,  \& {Hook}, R. 1997, The Tiny Tim User's Guide, Version 4.4
  (Baltimore: STScI)

\bibitem[\protect\citeauthoryear{Kritsuk}{Kritsuk}{1983}]{Kritsuk.1983.Afz.19.%
471}
Kritsuk, A.~G. 1983, Astrofizika, 19, 471

\bibitem[\protect\citeauthoryear{{Lavezzi} \& {Dickey}}{{Lavezzi} \&
  {Dickey}}{1998}]{Lavezzi.1998.AJ.115.405}
{Lavezzi}, T.~E.,  \& {Dickey}, J.~M. 1998, AJ, 115, 405

\bibitem[\protect\citeauthoryear{{McNamara}}{{McNamara}}{1997}]{McNamara.1997.%
GCCF.109}
{McNamara}, B.~R. 1997, in Galactic and Cluster Cooling Flows, ed. N.~Soker
  (San Francisco: ASP Press), 109

\bibitem[\protect\citeauthoryear{{McNamara} \& {Jaffe}}{{McNamara} \&
  {Jaffe}}{1994}]{McNamara.1994.AA.281.673}
{McNamara}, B.~R.,  \& {Jaffe}, W. 1994, A\&A, 281, 673

\bibitem[\protect\citeauthoryear{{McNamara} \& {O'Connell}}{{McNamara} \&
  {O'Connell}}{1989}]{McNamara.1989.AJ.98.2018}
{McNamara}, B.~R.,  \& {O'Connell}, R.~W. 1989, AJ, 98, 2018

\bibitem[\protect\citeauthoryear{{McNamara} \& {O'Connell}}{{McNamara} \&
  {O'Connell}}{1992}]{McNamara.1992.ApJ.393.579}
{McNamara}, B.~R.,  \& {O'Connell}, R.~W. 1992, ApJ, 393, 579

\bibitem[\protect\citeauthoryear{{McNamara} \& {O'Connell}}{{McNamara} \&
  {O'Connell}}{1993}]{McNamara.1993.AJ.105.417}
{McNamara}, B.~R.,  \& {O'Connell}, R.~W. 1993, AJ, 105, 417

\bibitem[\protect\citeauthoryear{{McNamara et al.}}{{McNamara et
  al.}}{1998}]{McNamara.1998.prep}
{McNamara et al.} 1998, in preparation

\bibitem[\protect\citeauthoryear{{Mihos} \& {Hernquist}}{{Mihos} \&
  {Hernquist}}{1994a}]{Mihos.1994.ApJ.425.L13}
{Mihos}, J.~C.,  \& {Hernquist}, L. 1994a, ApJ, 425, L13

\bibitem[\protect\citeauthoryear{{Mihos} \& {Hernquist}}{{Mihos} \&
  {Hernquist}}{1994b}]{Mihos.1994.ApJ.437.611}
{Mihos}, J.~C.,  \& {Hernquist}, L. 1994b, ApJ, 437, 611

\bibitem[\protect\citeauthoryear{{Mihos} \& {Hernquist}}{{Mihos} \&
  {Hernquist}}{1996}]{Mihos.1996.ApJ.464.641}
{Mihos}, J.~C.,  \& {Hernquist}, L. 1996, ApJ, 464, 641

\bibitem[\protect\citeauthoryear{{Nieto} et~al.}{{Nieto}
  et~al.}{1992}]{Nieto.1992.AA.257.97}
{Nieto}, J.-L., {Bender}, R., {Poulain}, P.,  \& {Surma}, P. 1992, A\&A, 257,
  97

\bibitem[\protect\citeauthoryear{Nulsen}{Nulsen}{1982}]{Nulsen.1982.MNRAS.198.%
1007}
Nulsen, P. E.~J. 1982, MNRAS, 198, 1007

\bibitem[\protect\citeauthoryear{{O'Connell} \& {McNamara}}{{O'Connell} \&
  {McNamara}}{1989}]{OConnell.1989.AJ.98.180}
{O'Connell}, R.~W.,  \& {McNamara}, B.~R. 1989, AJ, 98, 180

\bibitem[\protect\citeauthoryear{{O'Dea}, {Baum}, \& {Gallimore}}{{O'Dea}
  et~al.}{1994}]{ODea.1994.ApJ.436.669}
{O'Dea}, C.~P., {Baum}, S.~A.,  \& {Gallimore}, J.~F. 1994, ApJ, 436, 669

\bibitem[\protect\citeauthoryear{{Owen}, {Ledlow}, \& {Keel}}{{Owen}
  et~al.}{1995}]{Owen.1995.AJ.109.14}
{Owen}, F.~N., {Ledlow}, M.~J.,  \& {Keel}, W.~C. 1995, AJ, 109, 14

\bibitem[\protect\citeauthoryear{{Owen}, {White}, \& {Burns}}{{Owen}
  et~al.}{1992}]{Owen.1992.ApJS.80.501}
{Owen}, F.~N., {White}, R.~A.,  \& {Burns}, J.~O. 1992, ApJS, 80, 501

\bibitem[\protect\citeauthoryear{{Rees}}{{Rees}}{1989}]{Rees.1989.MNRAS.239.1P}
{Rees}, M.~J. 1989, MNRAS, 239, 1P

\bibitem[\protect\citeauthoryear{{Romanishin}}{{Romanishin}}{1987}]{Romanishin%
.1987.ApJ.323.L113}
{Romanishin}, W. 1987, ApJ, 323, L113

\bibitem[\protect\citeauthoryear{{Romanishin} \& {Hintzen}}{{Romanishin} \&
  {Hintzen}}{1988}]{Romanishin.1988.ApJ.324.L17}
{Romanishin}, W.,  \& {Hintzen}, P. 1988, ApJ, 324, L17

\bibitem[\protect\citeauthoryear{{Salpeter}}{{Salpeter}}{1955}]{Salpeter.1955.%
ApJ.121.161}
{Salpeter}, E.~E. 1955, ApJ, 121, 161

\bibitem[\protect\citeauthoryear{{Sarazin}}{{Sarazin}}{1986}]{Sarazin.1986.Rev%
ModPh.58.1}
{Sarazin}, C.~L. 1986, Rev. Mod. Phys., 58, 1

\bibitem[\protect\citeauthoryear{{Sarazin} et~al.}{{Sarazin}
  et~al.}{1995}]{Sarazin.1995.ApJ.447.559}
{Sarazin}, C.~L., {Burns}, J.~O., {Roettiger}, K.,  \& {McNamara}, B.~R. 1995,
  ApJ, 447, 559

\bibitem[\protect\citeauthoryear{{Sarazin} \& {McNamara}}{{Sarazin} \&
  {McNamara}}{1997}]{Sarazin.1997.ApJ.480.203}
{Sarazin}, C.~L.,  \& {McNamara}, B.~R. 1997, ApJ, 480, 203

\bibitem[\protect\citeauthoryear{{Scalo}}{{Scalo}}{1986}]{Scalo.1986.FCP.11.1}
{Scalo}, J.~M. 1986, Fundam. Cosmic Phys., 11, 1

\bibitem[\protect\citeauthoryear{{Smith} \& {Heckman}}{{Smith} \&
  {Heckman}}{1989}]{Smith.1989.ApJ.341.658}
{Smith}, R.~M.,  \& {Heckman}, T.~M. 1989, ApJ, 341, 658

\bibitem[\protect\citeauthoryear{{Sparks}, {Carollo}, \& {Macchetto}}{{Sparks}
  et~al.}{1997}]{Sparks.1997.ApJ.486.253}
{Sparks}, W.~B., {Carollo}, C.~M.,  \& {Macchetto}, F. 1997, ApJ, 486, 253

\bibitem[\protect\citeauthoryear{{Strom} \& {Strom}}{{Strom} \&
  {Strom}}{1979}]{Strom.1979.AJ.84.1091}
{Strom}, S.~E.,  \& {Strom}, K.~M. 1979, AJ, 84, 1091

\bibitem[\protect\citeauthoryear{{Sutherland}, {Bicknell}, \&
  {Dopita}}{{Sutherland} et~al.}{1993}]{Sutherland.1993.ApJ.414.510}
{Sutherland}, R.~S., {Bicknell}, G.~V.,  \& {Dopita}, M.~A. 1993, ApJ, 414, 510

\bibitem[\protect\citeauthoryear{{Taylor} et~al.}{{Taylor}
  et~al.}{1999}]{Taylor.1999.ApJ.512.L27}
{Taylor}, G.~B., {O'Dea}, C.~P., {Peck}, A.~B.,  \& {Koekemoer}, A.~M. 1999,
  ApJ, 512, L27

\bibitem[\protect\citeauthoryear{{Thomas} \& {Fabian}}{{Thomas} \&
  {Fabian}}{1990}]{Thomas.1990.MNRAS.246.156}
{Thomas}, P.~A.,  \& {Fabian}, A.~C. 1990, MNRAS, 246, 156

\bibitem[\protect\citeauthoryear{{Toomre} \& {Toomre}}{{Toomre} \&
  {Toomre}}{1972}]{Toomre.1972.ApJ.178.623}
{Toomre}, A.,  \& {Toomre}, J. 1972, ApJ, 178, 623

\bibitem[\protect\citeauthoryear{{Valluri} \& {Jog}}{{Valluri} \&
  {Jog}}{1990}]{Valluri.1990.ApJ.357.367}
{Valluri}, M.,  \& {Jog}, C.~J. 1990, ApJ, 357, 367

\bibitem[\protect\citeauthoryear{{Voit} \& {Donahue}}{{Voit} \&
  {Donahue}}{1997}]{Voit.1997.ApJ.486.242}
{Voit}, G.~M.,  \& {Donahue}, M. 1997, ApJ, 486, 242

\bibitem[\protect\citeauthoryear{{Walker}, {Mihos}, \& {Hernquist}}{{Walker}
  et~al.}{1996}]{Walker.1996.ApJ.460.121}
{Walker}, I.~R., {Mihos}, J.~C.,  \& {Hernquist}, L. 1996, ApJ, 460, 121

\bibitem[\protect\citeauthoryear{{Weil} \& {Hernquist}}{{Weil} \&
  {Hernquist}}{1993}]{Weil.1993.ApJ.405.142}
{Weil}, M.~L.,  \& {Hernquist}, L. 1993, ApJ, 405, 142

\end{thebibliography}

\clearpage

\begin{figure}
\figcaption[fig1.ps]
{Broad-band F702W \HST\ WFPC2 Planetary Camera image of A2597. The spatial
scale is $\sim 0\farcs0455$/pixel, corresponding to $\sim 65$~pc~/ pixel for
our adopted cosmological parameters. The orientation of the image is such that
north is to the top and east is to the left, and the original pointing
orientation of the image has been preserved (in order to avoid smoothing the
image by rotating the pixels).%
\label{fig:f702W}}
\end{figure}

\clearpage

\clearpage

\begin{figure}
\figcaption[fig3.ps]
{Broad-band $m_{\rm F450W} - m_{\rm F702W}$ color distribution of the galaxy,
obtained from the PC1 F702W and F450W images, each of which was convolved to
the resolution of the other before converting to magnitudes and performing the
subtraction. The magnitude units have been retained in the \HST\ VEGAMAG
system. We superpose contours from the 3.5~cm VLA radio continuum image 
	\protect\shortcite{Sarazin.1995.ApJ.447.559}.%
\label{fig:m450W-m702W}}
\end{figure}

\clearpage

\clearpage

\clearpage

\begin{figure}
\figcaption[fig6.ps]
{The greyscale represents the ``residual'' filamentary structure in the F702W
band, obtained by subtracting the elliptical galaxy model from the original
image. The contours are the 3.5~cm VLA radio continuum image 
	\protect\shortcite{Sarazin.1995.ApJ.447.559}.%
\label{fig:f702W_res}}
\end{figure}

\end{document}